\documentclass[twocolumn,english]{revtex4-1}
\usepackage[T1]{fontenc}
\usepackage[latin9]{inputenc}
\usepackage{color}
\usepackage{amsmath}
\usepackage{graphicx}

\makeatletter

\newcommand{\lyxdot}{.}

\usepackage{hyperref}

\makeatother

\usepackage{babel}
\begin{document}
\title{Noise-aware neural network for stochastic dynamics simulation }
\author{Pei-Fang Wu}
\thanks{These two authors contributed equally to this work.}
\affiliation{Institute for Theoretical Physics, School of Physics, South China
Normal University, Guangzhou 510006, China}
\affiliation{Key Laboratory of Atomic and Subatomic Structure and Quantum Control
(Ministry of Education), School of Physics, South China Normal University,
Guangzhou 510006, China}
\affiliation{Guangdong Provincial Key Laboratory of Quantum Engineering and Quantum
Materials, South China Normal University, Guangzhou 510006, China}
\author{Wei-Chen Guo}
\thanks{These two authors contributed equally to this work.}
\affiliation{Institute for Theoretical Physics, School of Physics, South China
Normal University, Guangzhou 510006, China}
\affiliation{Key Laboratory of Atomic and Subatomic Structure and Quantum Control
(Ministry of Education), School of Physics, South China Normal University,
Guangzhou 510006, China}
\affiliation{Guangdong Provincial Key Laboratory of Quantum Engineering and Quantum
Materials, South China Normal University, Guangzhou 510006, China}
\author{Liang He}
\email{liang.he@scnu.edu.cn}

\affiliation{Institute for Theoretical Physics, School of Physics, South China
Normal University, Guangzhou 510006, China}
\affiliation{Key Laboratory of Atomic and Subatomic Structure and Quantum Control
(Ministry of Education), School of Physics, South China Normal University,
Guangzhou 510006, China}
\affiliation{Guangdong Provincial Key Laboratory of Quantum Engineering and Quantum
Materials, South China Normal University, Guangzhou 510006, China}
\begin{abstract}
In the presence of system-environment coupling, classical complex
systems undergo stochastic dynamics, where rich phenomena can emerge
at large spatio-temporal scales. To investigate these phenomena, numerical
approaches for simulating stochastic dynamics are indispensable and
can be computationally expensive. In light of the recent fast development
in machine learning techniques,\textcolor{red}{{} }here, we establish
a generic machine learning approach to simulate the stochastic dynamics,
dubbed the noise-aware neural network (NANN). One key feature of this
approach is its ability to generate the long-time stochastic dynamics
of complex large-scale systems by just training NANN with the one-step
dynamics of smaller-scale systems, thus reducing the computational
cost. Furthermore, this NANN based approach is quite generic. Case-by-case
special design of the architecture of NANN is not necessary when it
is employed to investigate different stochastic complex systems. Using
the noisy Kuramoto model and the Vicsek model as concrete examples,
we demonstrate its capability in simulating stochastic dynamics. We
believe that this novel machine learning approach can be a useful
tool in investigating the large spatio-temporal scaling behavior of
complex systems subjected to the influences of the environmental noise.
\end{abstract}
\maketitle

\section{Introduction}

Recently, employing machine learning techniques to boost molecular
dynamics simulation has aroused increasing attention \citep{Farimani_JCP_2022,behler_PRL_2007,Chmiela_SciAdv_2017,Zhanglinfeng_PRL_2018,Zhanglinfeng_MP_2019,Zhanglinfeng_JCP_2021,Zhanglinfeng_CPC_2021,Zhongzhicheng_PRB_2022,Wanglei_PRX_2020,Wanglei_JML_2022,Wanglei_SciPost_2023}.
Various approaches have been developed and successfully applied. These
include the graph neural network accelerated molecular dynamics \citep{Farimani_JCP_2022},
the Behler-Parrinello neural network \citep{behler_PRL_2007}, the
gradient-domain machine learning \citep{Chmiela_SciAdv_2017}, the
deep potential method \citep{Zhanglinfeng_PRL_2018,Zhanglinfeng_MP_2019,Zhanglinfeng_JCP_2021,Zhanglinfeng_CPC_2021,Zhongzhicheng_PRB_2022},
the neural canonical transformation \citep{Wanglei_PRX_2020,Wanglei_JML_2022,Wanglei_SciPost_2023},
etc. However, a closely related type of dynamics, namely, stochastic
dynamics \citep{honerkamp_book_1996}, has received much less attention
in this context so far \citep{Casert_arXiv_2022,MKJ_arXiv_2023,Jeong_SciRep_2021,Zhu_JCP_2023}.
In fact, due to the ubiquitous system-environment coupling, stochastic
dynamics is one fundamental type of dynamics appearing in many fields,
ranging from active matter \citep{Solon_PRL_2015,Omar_PRL_2021,wysocki_EuroLett_2014,siebert_Soft_2017,Digregorio_PRL_2018,Speck_PRE_2021,fernandez_NatCom_2020,rein_EuroLett_2023,Vicsek_PRL_1995,Chate_PRL_2004,Toner_Ann_Phys_2005,Chate_PRE_2008},
over intelligent matter \citep{Boguna_PRE_2015,Boguna_PRE_2017,Yusufaly_PRE_2016_QuorumSensing,Aguilar_PRE_2021_QuorumSensing,Tobias_Nat_Commun_2018_QuorumSensing,Lowen_PNAS_2021},
to social physics \citep{farkas_physA_2003}, etc. Frequently, rich
phenomena can emerge at large spatio-temporal scales in classical
complex systems undergoing stochastic dynamics \citep{kursten_PRL_2020,ginelli_PRL_2010,sumino_Nat_2012}.
This naturally places demanding tasks for numerical approaches to
simulate stochastic dynamics.

Conventionally, various strategies can developed for accelerating
such stochastic dynamics simulations in certain cases to meet the
above demand, however, they are often developed on a case-by-case
basis. Noticing that the current applications of machine learning
techniques in physics \citep{Carleo_RMP_2019,Carrasquilla_AdvPhysX_2020,Volpe_Nat_Mach_Intell_2020}
widely make use of the universality and transferability of machine
learning models, especially the applications of artificial neural
networks \citep{Melko_Nat_Phys_2017,van_Nieuwenburg_Nat_Phys_2017,Wanglei_PRL_2018_RG,Guo_NJP_2023},
this thus raises the intriguing question of whether and how neural
networks can be developed as a generic tool for stochastic dynamics
simulation. For this purpose, a natural approach to accelerating stochastic
dynamics simulations with neural networks is through the extrapolation
of the system size \citep{Farimani_JCP_2022,behler_PRL_2007,Chmiela_SciAdv_2017,Zhanglinfeng_PRL_2018,Zhanglinfeng_MP_2019,Zhanglinfeng_JCP_2021,Zhanglinfeng_CPC_2021,Zhongzhicheng_PRB_2022,Wanglei_PRX_2020,Wanglei_JML_2022,Wanglei_SciPost_2023},
i.e., generating datasets using conventional stochastic dynamics simulations
on small system sizes, training neural networks on these datasets,
and asking neural networks to generate the stochastic dynamics at
larger system sizes. However, this imposes higher demands on the neural
networks in addition to the difficulties of extrapolation itself,
as they are asked to predict the phenomena that might not exist in
the small system sizes. Therefore, the neural networks shall not merely
learn specific patterns, configurations, or evolutionary trajectories,
but shall also learn the underlying logic of the physical model, namely
learn the stochastic dynamical equations themselves. 

\begin{figure}
\begin{centering}
\includegraphics[width=3.3in]{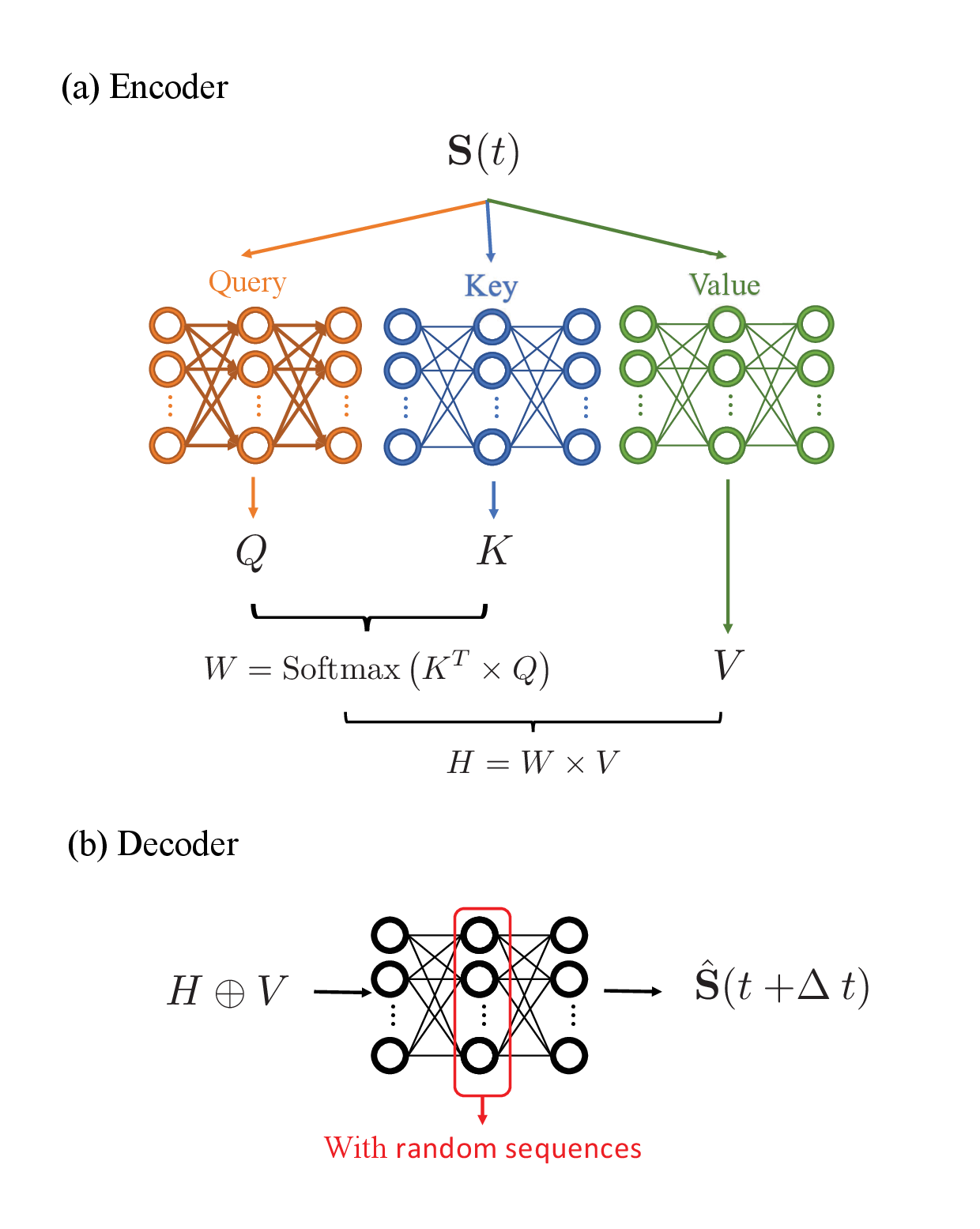}
\par\end{centering}
\caption{\textcolor{blue}{\label{NANN}}Schematic illustration of NANN's architecture,
which includes (a) encoder and (b) decoder. The encoder encodes the
input $\mathbf{S}(t)$ into key $K$, query $Q$, and value $V$,
leading to an attention weight $W$ and further to a state latent
variable $H$. This compresses the input data sequence into a latent
space representation while capturing intricate relationships within
the sequence. The state latent variable $H$ can then be decoded by
the decoder, where random sequences are introduced into every fully-connected
layer to match the influences from stochastic noise. After training
via minimizing the moment loss function $l$ that estimates the difference
between the predicted state $\hat{\mathbf{S}}(t+\Delta t)$ and the
target $\mathbf{S}(t+\Delta t)$, the predicted state $\hat{\mathbf{S}}(t+\Delta t)$
is expected to assume a similar probability distribution as the one
obtained directly by the simulations. It is demonstrated in this work
that NANN can learn to generate the long-term stochastic dynamics
of the relatively larger system sizes by just training with the one-step
dynamics on the relatively smaller system sizes. See text for more
details.}
\end{figure}

In this work, we address this question by proposing a new type of
neural network, dubbed noise-aware neural network (NANN) (see Fig.~\ref{NANN}),
based on which a generic machine learning approach is developed to
simulate stochastic dynamics. This approach assumes two key features:
(I) It is able to generate the long-time stochastic dynamics of complex
large-scale systems by just training NANN with the one-step dynamics
of smaller-scale systems. (II) Case-by-case special design of the
architecture of NANN is not necessary when employed to investigate
different stochastic complex systems. 

\begin{figure}
\begin{centering}
\textcolor{black}{\includegraphics[width=1.65in]{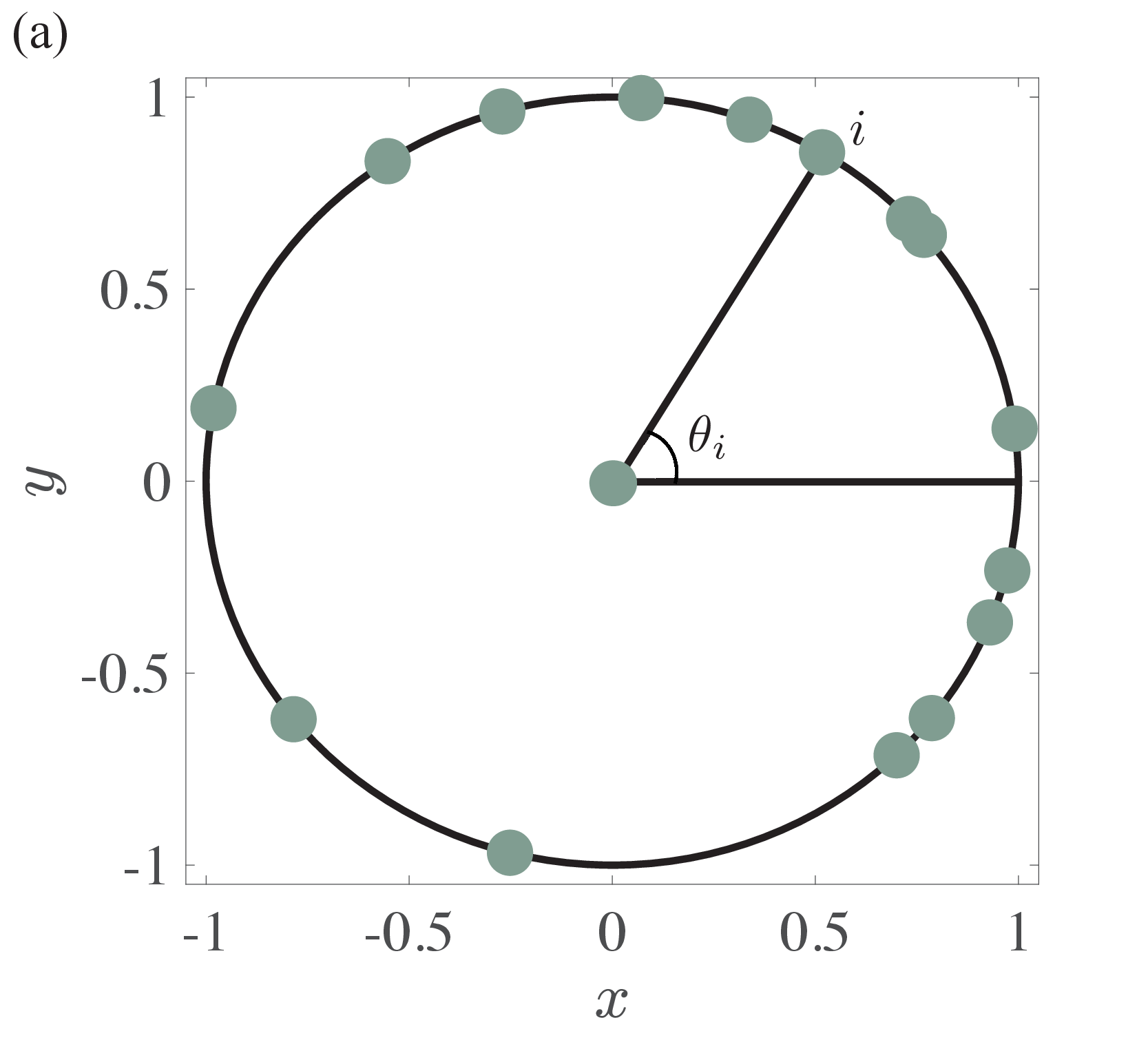}\includegraphics[width=1.65in]{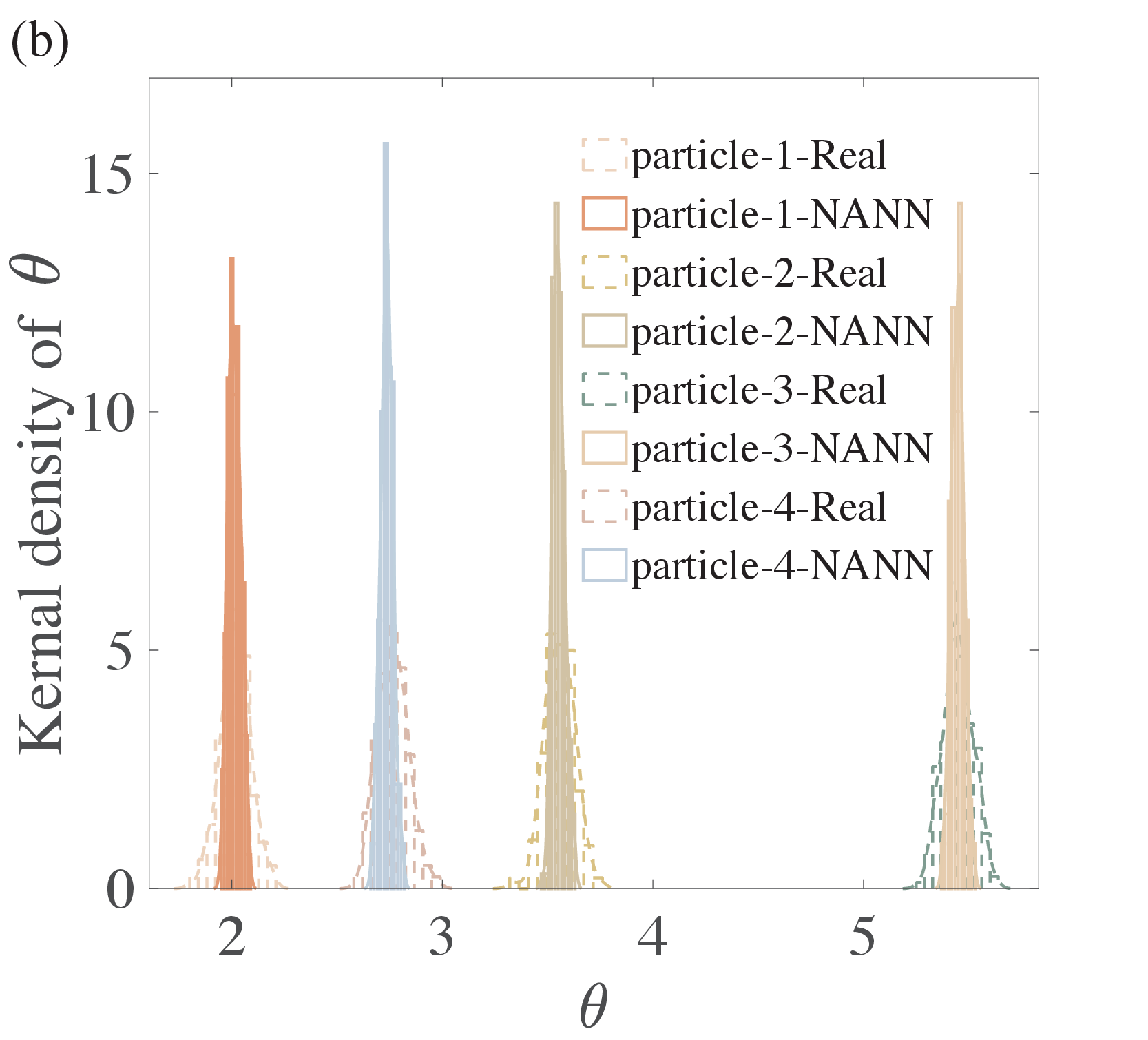}}
\par\end{centering}
\begin{centering}
\textcolor{black}{\includegraphics[width=1.65in]{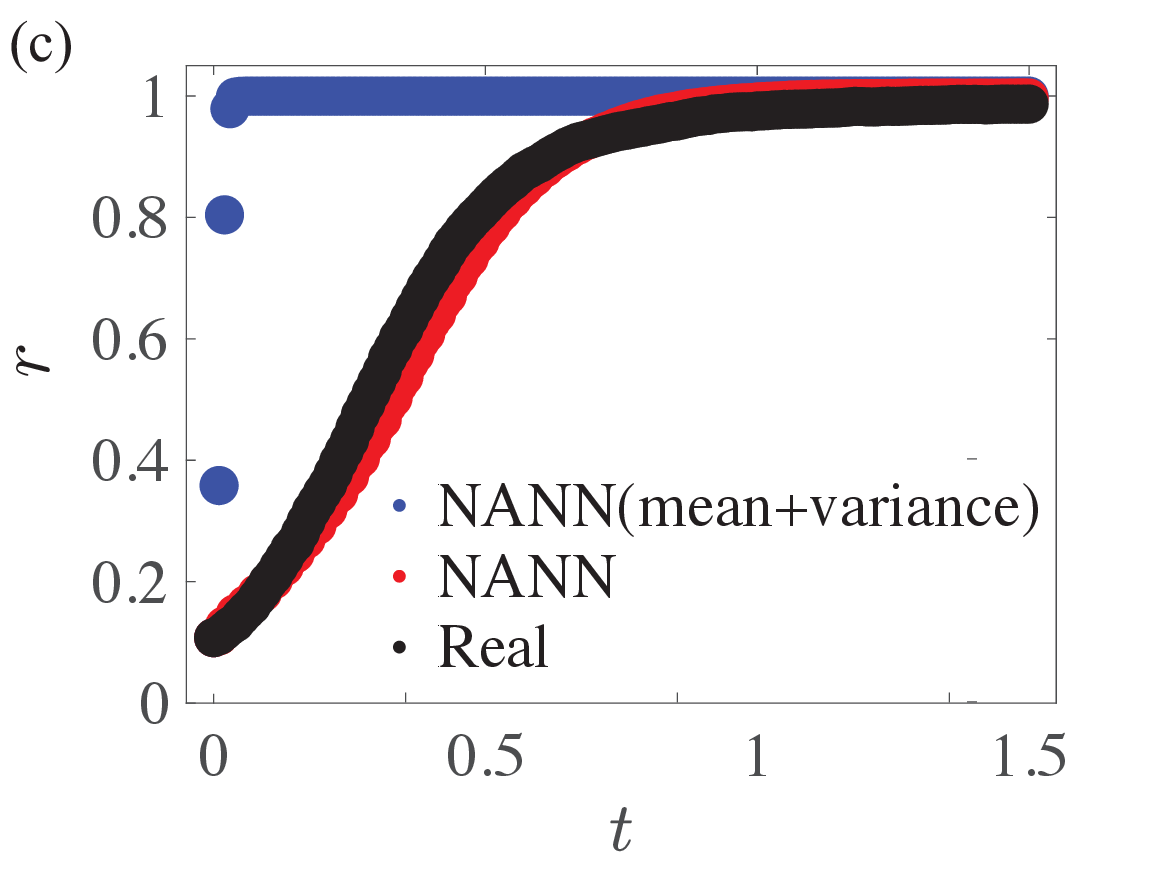}\includegraphics[width=1.65in]{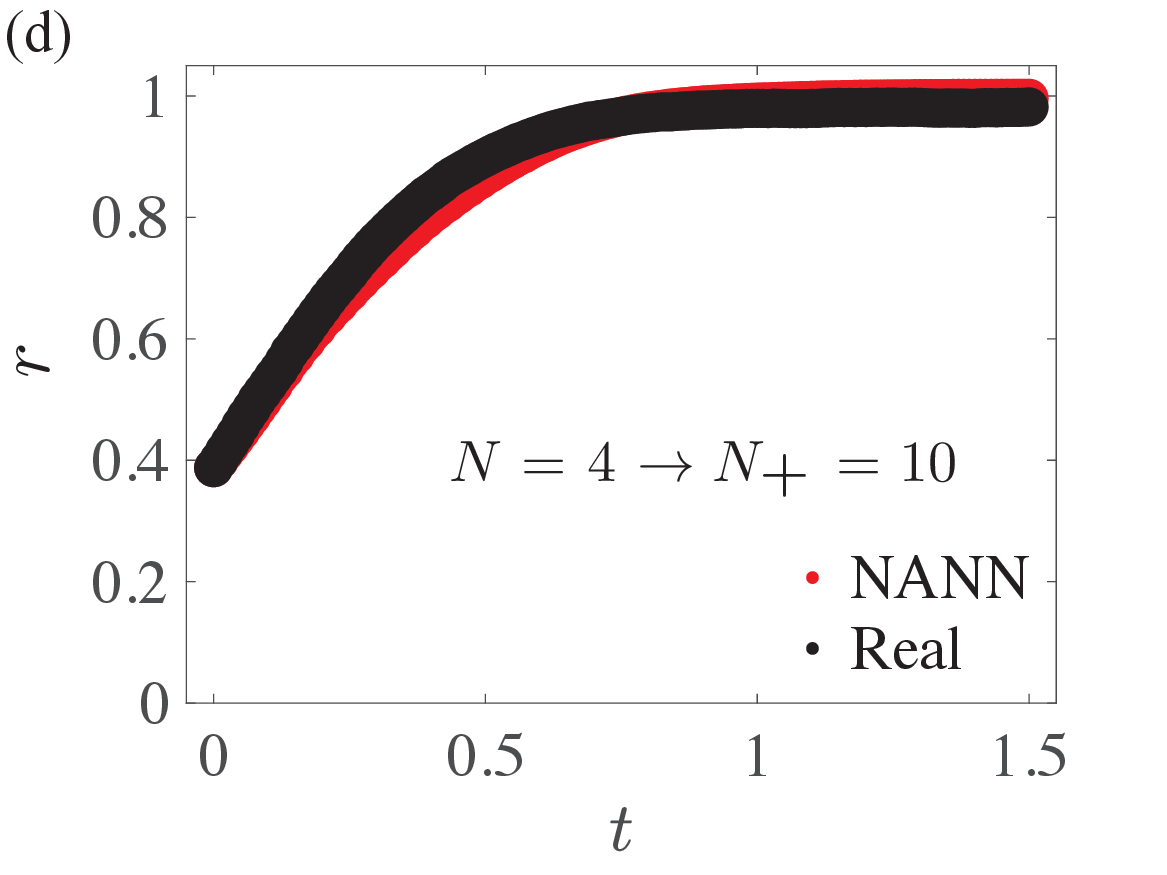}}
\par\end{centering}
\begin{centering}
\textcolor{black}{\includegraphics[width=1.65in]{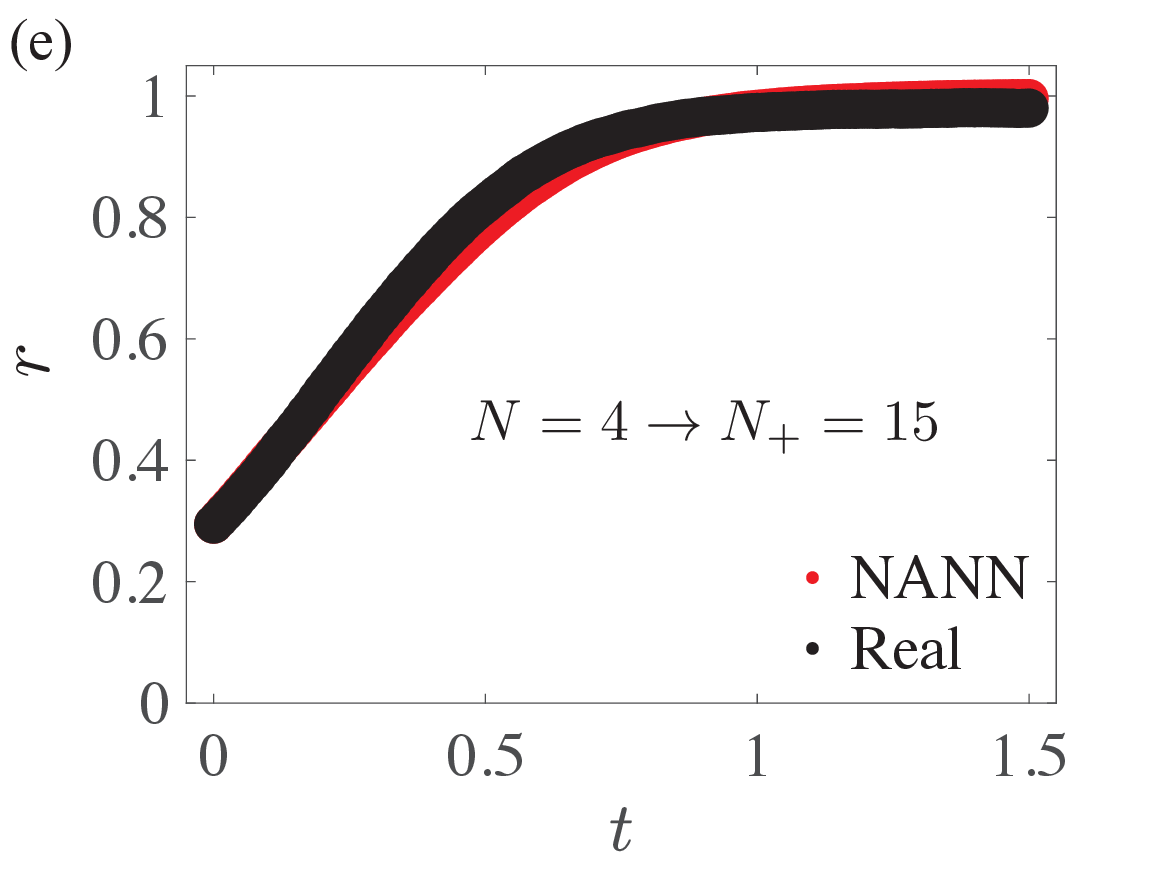}\includegraphics[width=1.65in]{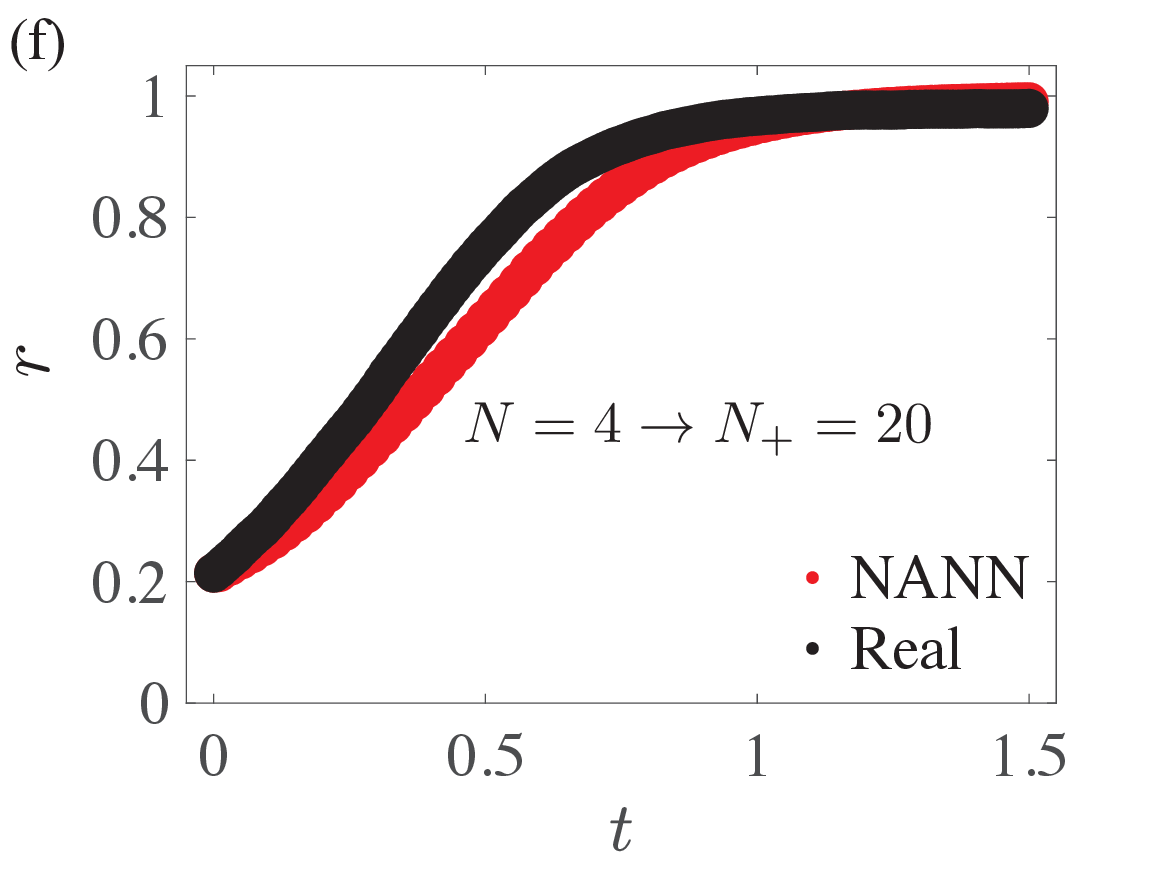}}
\par\end{centering}
\textcolor{black}{\caption{\textcolor{blue}{\label{Kuramoto}}NANN based simulations on the noisy
Kuramoto model. (a) Schematic illustration of the phase $\theta$
distribution of the noisy Kuramoto model consisting of $15$ limit-cycle
oscillators for instance. (b) Direct comparison of the probability
distribution of every oscillator's phase in the subsequent time step
concerning the noisy Kuramoto model's one-step dynamics, with the
solid lines corresponding to the NANN's outcomes and the dashed lines
corresponding to the results of the real simulations on this noisy
Kuramoto model with $N=4$. (c) Direct comparison of the time evolution
of the order parameter $r$ concerning the noisy Kuramoto model's
long-term stochastic dynamics, with the red markers corresponding
to the NANN's outcomes with $A=4$, the blue markers corresponding
to the NANN's outcomes with $A=2$, and the black markers corresponding
to the results of the real simulations on this Vicsek model with $N=4$.
Given only the $t=0$ one-step data for training, when only the first-order
and second-order moments are taken into account in the moment loss
function $l$, i.e., when $A=2$, the machine learning is completely
unable to capture the physics. While in sharp contrast, the NANN trained
with $A=4$ accomplishes its task very well. Furthermore, it successfully
generates the long-term dynamics of the (d) $N_{+}=10$, (e) $N_{+}=15$,
(f) $N_{+}=20$ systems by just training on $N=4$. See text for more
details.}
}
\end{figure}

We demonstrate the capability of this approach in two prototypical
models, namely, noisy Kuramoto model \citep{Acebrn_RevOfModernPhys_2005,rodrigues_PhysRep_2016}
and Vicsek model \citep{Vicsek_PRL_1995,Toner_Ann_Phys_2005,Chate_PRL_2004,Chate_PRE_2008}.
More specifically, in the former case, by training NANN to analyze
the one-step evolution of the phase distribution of 4 oscillators,
NANN successfully predicts the long-time dynamics of not only 4, but
$10,15,20$ oscillators as well (see Fig.~\ref{Kuramoto}). The consistency
between NANN's extrapolating outputs and the results of the conventional
simulations clearly manifests the key feature (I) of NANN. Compared
with many neural network-based approaches each instance of the network
is specifically trained for a fixed system size, this clearly manifests
NANN's potential to boost the stochastic dynamics simulation. In the
latter case of the Vicsek model, although it assumes a completely
different type of interaction compared with the one of the noisy Kuramoto
model, we adopt the same network architecture for the NANN and successfully
generate the long-time dynamics for a relatively large system by training
NANN with the one-step dynamics of a smaller system (see Fig.~\ref{Vicsek}),
clearly manifesting the key features (I) and (II) of NANN. Moreover,
the efficiency of using NANN to generate evolution trajectories can
be about one order of magnitude higher than that of using the conventional
algorithm to simulate the stochastic dynamics at the same system size.
These findings suggest that NANN is a promising generic tool that
can be readily applied to investigate stochastic dynamics of various
complex systems. 

\section{Noise-aware neural network based approach for stochastic dynamics
simulations}

For a many-body complex system consisting of $N$ particles, when
$M$ state variables (e.g., the particle's position and velocity)
are under consideration, the stochastic dynamics simulations \citep{honerkamp_book_1996}
aim at finding the system's state $\mathbf{S}(t)\equiv\{S_{i}(t)\}$
($i=1,2,\ldots,N$) at any time $t$, with $S_{i}(t)\equiv\{s_{i,m}(t)\}$
($m=1,2,\ldots,M$) describing the $i$th particle's state, and $s_{i,m}(t)$
is the value of the $m$th state variable for the $i$th particle.
To realize it with the assistance of machine learning, the first concrete
goal is to train a machine learning model with a series of known $\mathbf{S}(t)$
obtained directly by the simulations, and then let the machine learning
model predict the system's state in the subsequent time step $t+\Delta t$,
where $\Delta t$ is an arbitrary time step. After training, this
predicted state $\hat{\mathbf{S}}(t+\Delta t)$ shall be as similar
as possible to the target $\mathbf{S}(t+\Delta t)$ to be predicted,
i.e., the real state in the subsequent time step obtained directly
by the simulations. The difference between the predicted state $\hat{\mathbf{S}}(t+\Delta t)$
and the target $\mathbf{S}(t+\Delta t)$ is estimated by a loss function
$l$. When $\hat{\mathbf{S}}(t+\Delta t)$ and $\mathbf{S}(t+\Delta t)$
are both deterministic, a prototypical choice of the loss function
is the mean square error (MSE) \citep{Goodfellow_Book_2016}. But
due to the influences from \textcolor{black}{stochastic} noise, here
the target $\mathbf{S}(t+\Delta t)$ always assumes probabilistic
changes. Normally the data that one can input to a machine learning
model, as well as the machine learning model's outcomes after processing
the data, are specific states of the system. This is one of the fundamental
challenges that make the machine learning prediction and acceleration
of stochastic dynamics simulations much more complex than those general
applications of machine learning techniques in molecular dynamics
simulations \citep{Farimani_JCP_2022,behler_PRL_2007,Chmiela_SciAdv_2017,Zhanglinfeng_PRL_2018,Zhanglinfeng_MP_2019,Zhanglinfeng_JCP_2021,Zhanglinfeng_CPC_2021,Zhongzhicheng_PRB_2022,Wanglei_PRX_2020,Wanglei_JML_2022,Wanglei_SciPost_2023}.
To overcome this challenge, we construct a moment loss function 
\begin{equation}
l=\sum_{m}^{M}\mathrm{MSE}\left(\sum_{a}^{A}\left\langle \hat{\mathbf{S}}_{m}^{a}(t)\right\rangle ,\sum_{a}^{A}\left\langle \mathbf{S}_{m}^{a}(t)\right\rangle \right),\label{loss}
\end{equation}
where $a$ is the order of the moment, and $A$ is the highest order
that is taken into account for machine learning. During the training,
this moment loss function $l$ is minimized by an optimization algorithm
(e.g., the Adam algorithm \citep{Kingma_ICLR_2015}), which is expected
to promote the machine learning model's outcomes to meet the underlying
logic of the physical model, hence making it possible to capture the
influences from \textcolor{black}{stochastic} noise. But this is still
far from all one needs for predicting and accelerating stochastic
dynamics simulations.

The extrapolation of system size is another challenge that couples
with the challenge of the \textcolor{black}{stochastic} noise. If
a machine learning model can predict stochastic dynamics on relatively
larger system sizes (assuming $N_{+}$ particles) by just training
on the relatively smaller ones (assuming $N<N_{+}$ particles), it
is then possible to accelerate the simulations in practice. For this
purpose, before asking whether the stochastic dynamics of different
system sizes can be generated effectively and efficiently, the machine
learning model is required to be capable, at least in form, of flexibly
dealing with multi system sizes. Ordinary feedforward architectures
are not naturally suitable in this context \citep{Goodfellow_Book_2016},
and we shall focus on the network architectures with attention \citep{vaswani_Advances_2017}.
The so-called attention is the key to the famous transformer architectures,
which is also increasingly being applied to assisting physical investigations
\citep{alkin_arxiv_2024,geneva_NN_2022,xu_Fluids_2024,wu_arxiv_2024}.
It generally enables the machine learning model to flexibly deal with
multi system sizes.

Incorporating these, we establish the noise-aware neural network (NANN)
that combines a self-attention block \citep{vaswani_Advances_2017}
within the encoder \citep{sutskever_Advances_2014}, enabling the
compression of the input data sequence into a latent space representation
while capturing intricate relationships within the sequence. The decoder
is composed of several fully-connected layers \citep{Goodfellow_Book_2016},
augmented by introduced random sequences. Such a setup can produce
distinct outputs for the same input, corresponding to the variations
observed in relevant systems due to stochastic noise. The PReLU function
$\textrm{PReLU}(x)\equiv\max(0,x)+a\min(0,x)$ (where $a$ is the
learnable parameter) \citep{he_IEEE_2015} is applied as the activation
function \citep{Goodfellow_Book_2016} since it allows the effective
prediction of the distribution of state variables containing negative
values. Each input sample for NANN is a certain state $\mathbf{S}(t)$
that includes all the $M$ state variables of all the $N$ particles
of the system at time $t$, which is a tensor with dimension $\left[N,M\right]$.
The data samples are first encoded through a fully-connected layer
$\boldsymbol{x}\equiv\left(x_{1},x_{2},\cdots,x_{M}\right)^{T}$ with
$M$ neurons, followed by passing the encoded input samples to the
attention block in the encoder. In the attention block, the key $K$
and query $Q$ contain information that determines the strength of
interactions of each particle in the system with respect to all other
particles, thus outputting an attention weight $W$ with dimension
$\left[N,N\right]$ that models the strength of interactions among
all particles. The value $V$ retains the characteristics of the input
samples and is intended to be combined with the weight matrix to update
the state representation of the latent space of the particle. Eventually,
the updated state representation of particles in the latent space
is decoded by a decoder consisting of fully-connected layers containing
random sequences, so as to obtain the predicted state $\hat{\mathbf{S}}(t+\Delta t)$.

\textcolor{black}{}

\section{NANN based simulation of the noisy Kuramoto model}

To illustrate concretely how NANN can be applied in simulating stochastic
dynamics, let us start with the noisy Kuramoto model \citep{Acebrn_RevOfModernPhys_2005,rodrigues_PhysRep_2016}
as a preliminary demonstration. The physical system under consideration
consists of $N$ limit-cycle oscillators with the presence of stochastic
noise, whose dynamical equation reads \textcolor{black}{\citep{Acebrn_RevOfModernPhys_2005,rodrigues_PhysRep_2016}}

\textcolor{black}{
\begin{equation}
\dot{\theta_{i}}\left(t\right)=\omega_{i}+\eta_{i}\left(t\right)+\frac{J}{N}\sum_{j=1}^{N}\sin\left(\theta_{j}\left(t\right)-\theta_{i}\left(t\right)\right).\label{eq:Kuramoto}
\end{equation}
Here, }$\eta_{i}(t)$ is a stochastic Gaussian noise,\textcolor{black}{{}
$\theta_{i}\left(t\right)$ is the phase of the $i$th oscillator,
$\omega_{i}$ is its intrinsic natural frequency, and $J$ is a coupling
strength. The }phase coherence $r\equiv\langle\exp(i\theta(t))\rangle$
of all oscillators is a typical order parameter in this system, which
reaches its maximum $r=1$ when all oscillators' phases are \textcolor{black}{synchronous,
and its minimum $r=0$ when the phases are balanced around the circle,
such as evenly spread or in clusters that balance each other out \citep{Acebrn_RevOfModernPhys_2005,rodrigues_PhysRep_2016}.
With all oscillators constrained to lie on the unit circle as illustrated
in Fig.~}\ref{Kuramoto}\textcolor{black}{(a), }the \textcolor{black}{positions
of every oscillator are determined by $x_{i}(t)\equiv\cos\theta_{i}(t),y_{i}(t)\equiv\sin\theta_{i}(t)$,
and hence the phase $\theta_{i}\left(t\right)$ is the only }state
variable under consideration, i.e., $S_{i}(t)=\{\theta_{i}(t)\}$.

\textcolor{black}{In this scenario},\textcolor{black}{{} we train NANN
with a series of known }$\mathbf{S}(t)=\{S_{i}(t)\}$ ($i=1,2,\ldots,N$),
and then let NANN predict the system's state at $t+\Delta t$. This
predicted state $\hat{\mathbf{S}}(t+\Delta t)$ is expected to hold
the physical meaning of the possible phase configuration of all $N$
\textcolor{black}{oscillators at $t$, and it shall be }as similar
as possible to the target $\mathbf{S}(t+\Delta t)$, i.e., the phase
configuration obtained directly by the simulations. However, due to
the target's probabilistic changes driven by the stochastic noise,
it is pointless to require the predicted phase for every $i$ and
$t$ to be exactly the same as the one obtained directly by the simulations.
Instead, when NANN learns the probability distribution that governs
the generation of the target, one knows that its outputs actually
simulate the dynamical equation under consideration, namely Eq.~(\ref{eq:Kuramoto}),
and thus the time evolution of the order parameter $r$ shall be consistent
between the one calculated using NANN's outcomes and the one calculated
using the results of the real simulations. We indeed find such results,
as we shall see in the following.

Here, we fix \textcolor{black}{$J=1$ and assume that every oscillator
has the same $\omega_{i}=0.3$ for instance. The system is initialized
to a typical state that the oscillator's phase spans from $0$ to
$2\pi$, and this initial state }is provided to train the NANN, with
any $t>0$ being a subsequent time step to be predicted. We first
confirm that NANN can indeed learn the probability distribution that
governs the generation of the target for the one-step evolution, as
shown in \textcolor{black}{Fig.~}\ref{Kuramoto}\textcolor{black}{(b).
Then we examine its performance concerning the long-term }dynamics.
Fig.~\ref{Kuramoto}(c) shows a direct comparison of the time evolution
of the order parameter $r$ for $N=4$ oscillators. The black curve
is the $r$ calculated using the results of the real simulations.
The blue curve is the $r$ calculated using NANN's outcomes with $A=2$,
i.e., it is trained with only the first-order moment (mean) and the
second-order moment (variance). The red curve is the $r$ calculated
using NANN's outcomes with $A=4$.

From the blue curve in Fig.~\ref{Kuramoto}(c), one can clearly see
that although the moment loss function $l$ provides a general framework
for learning probability distributions, the powerful fitting ability
of machine learning is completely unable to capture the physics when
only the first-order and second-order moments are taken into account.
This thus illustrates the technical difficulties of applying machine
learning techniques to predicting and even accelerating stochastic
dynamics simulations. Despite that, the NANN trained with $A=4$ predicts
the stochastic dynamics simulations of this noisy Kuramoto model very
well, as shown clearly in Fig.~\ref{Kuramoto}(c). The time evolution
of the order parameter $r$ is highly consistent, manifesting that
given only the $t=0$ one-step data for training, NANN has the effectiveness
to predict the system's state at arbitrary subsequent time steps.

\textcolor{black}{Furthermore, NANN does not fixedly correspond to
a specific particle number $N$, }so \textcolor{black}{it might have
an }extrapolation capability concerning the system size. Such an extrapolation
is no doubt a non-trivial task, and its effectiveness needs to be
examined in practice. But somewhat surprisingly, we find that NANN
can indeed realize it very well. Fig.~\ref{Kuramoto}(d-f) shows
the direct comparisons of the time evolution of the order parameter
$r$ concerning different $N_{+}$. The black curves are the $r$
calculated using the results of the real simulations on $N_{+}=10,15,20$
systems, respectively. And the red curves are the $r$ calculated
using NANN's outcomes with $A=4$, just training on a relatively smaller
$N=4$ system and then extrapolating to predict stochastic dynamics
on the relatively larger $N_{+}=10,15,20$ systems, respectively.
In a detailed check on the probability distributions of each particle,
we confirm that extrapolating system size $N\rightarrow N_{+}$ does
not break NANN's effectiveness in learning the probability distribution.
The \textcolor{black}{consistency between NANN's }extrapolating outputs
and the results of the real simulations is not only in the sense of
ensemble average, but works for every particle at the same time. These
results thus indicate that NANN can effectively deal with stochastic
dynamics simulations, overcoming not only the challenge from the \textcolor{black}{stochastic}
noise, but also the challenge from the extrapolation of system size.

\section{Application of NANN in the flocking dynamics}

In this section, let us switch to another prototypical but more complex
model, namely the Vicsek model \citep{Vicsek_PRL_1995,Toner_Ann_Phys_2005,Chate_PRL_2004,Chate_PRE_2008},
to better demonstrate the effectiveness, efficiency, and generality
of the NANN bashed approach in simulating stochastic dynamics. The
physical system under consideration consists of $N$ self-propelled
active particles in a two-dimensional box of size $L\times L$ with
the presence of extrinsic stochastic noise, whose dynamical equation
reads \citep{Vicsek_PRL_1995,Toner_Ann_Phys_2005,Chate_PRL_2004,Chate_PRE_2008}

\textcolor{black}{
\begin{align}
\boldsymbol{x}_{j}(t+\Delta t) & =\boldsymbol{x}_{j}(t)+\boldsymbol{v}_{j}(t)\Delta t,\label{eq:Position}\\
\theta_{j}(t+\Delta t) & =\arg\sum_{k\in U_{j}}(e^{i\theta_{k}(t)}+\eta e^{i\xi_{j}(t)}).\label{eq:Direction_with_extrinsic_noise}
\end{align}
Here, $\boldsymbol{x}_{j}(t)$, $\boldsymbol{v}_{j}(t)$ and $\theta_{j}(t)$
are the position, the velocity, and the direction of motion of the
$j$th particle at time $t$, respectively. $\xi_{j}(t)\in[-\pi,\pi]$
is a uniformly distributed random noise with $\eta\in[0,1]$ being
the extrinsic noise level. With all self-propelled particles assuming
the same constant speed $v_{0}$, each particle adjusts its direction
of motion in the time step based on the average velocity direction
of all particles within its neighbors $U_{j}$ within a circular region
of radius $R$. Concerning the flocking dynamics }in this system\textcolor{black}{,
the group velocity }$v_{a}=\vert\sum_{j=1}^{N}\boldsymbol{v}_{j}\vert\slash(Nv_{0})$
of all self-propelled particles is a typical order parameter \citep{Vicsek_PRL_1995,Toner_Ann_Phys_2005,Chate_PRL_2004,Chate_PRE_2008},
which reaches its maximum $v_{a}=1$ when the particles form a flock
with their directions of motion are all aligned, and its minimum $v_{a}=0$
when the directions of motion are totally random. Distinguishing from
the global interactions involved in the noisy Kuramoto model, the
alignment involved here in the Vicsek model acts locally, and is even
trickier since at the macroscopic level it is not Newtonian but actually
based on the intelligent communicating and decision-making of the
active agents \citep{Vicsek_PRL_1995,Toner_Ann_Phys_2005,Chate_PRL_2004,Chate_PRE_2008}.
Therefore, this is not only a typical scenario that can better demonstrate
NANN, but also a practical application scenario where \textcolor{black}{more
intricate collective phenomena are expected to emerge at larger scales
\citep{kursten_PRL_2020}}, calling for the assistance of machine
learning.

\textcolor{black}{}
\begin{figure}
\begin{centering}
\textcolor{black}{\includegraphics[width=1.65in]{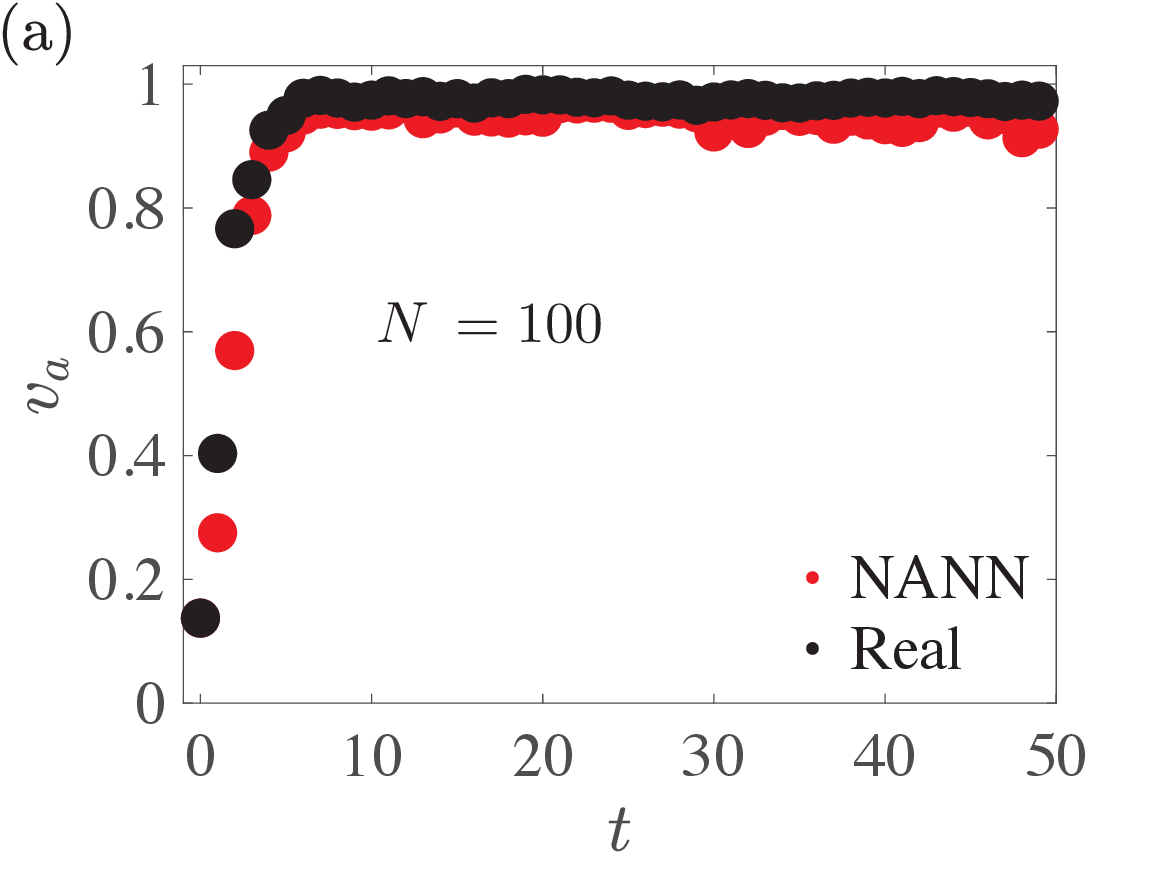}\includegraphics[width=1.65in]{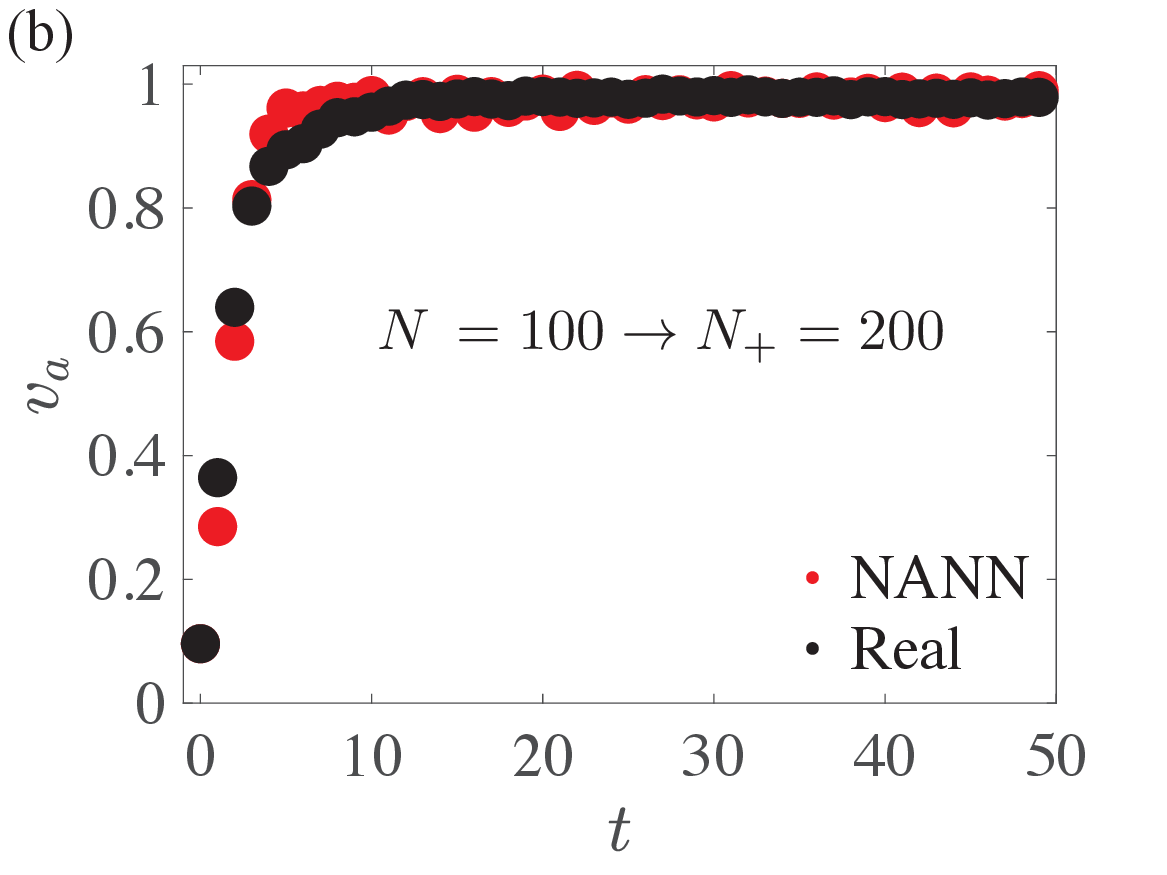}}
\par\end{centering}
\centering{}\textcolor{black}{\caption{\textcolor{blue}{\label{Vicsek}}NANN's application on simulations
of flocking dynamics. (a) Direct comparison of the time evolution
of the order parameter $v_{a}$ concerning the Vicsek model's long-term
flocking dynamics, with the red markers corresponding to the NANN's
outcomes and the black markers corresponding to the results of the
real simulations on this Vicsek model with $N=100$. The consistency
in $v_{a}$ manifests that, given only the $t=0$ one-step data for
training, NANN has the effectiveness to predict this nonequilibrium
complex system's long-term dynamics. (b) The extrapolation capability
of NANN in this scenario. NANN successfully generates the long-term
dynamics of the $N_{+}=200$ system by just training on $N=100$.
See text for more details.}
}
\end{figure}

\textcolor{black}{In this scenario, $\boldsymbol{x}_{j}(t)$ and $\theta_{j}(t)$
are the state variables under consideration, }i.e., $S_{j}(t)=\{\boldsymbol{x}_{j}(t),\theta_{j}(t)\}$.\textcolor{black}{{}
}Using the moment loss function $l$ with $A=4$\textcolor{black}{,
we train NANN with a series of known }$\mathbf{S}(t)=\{S_{j}(t)\}$
($j=1,2,\ldots,N$), and then let NANN predict the system's state
at $t+\Delta t$. This predicted state $\hat{\mathbf{S}}(t+\Delta t)$
is expected to hold the physical meaning of the possible position
and velocity configuration of all $N$ \textcolor{black}{self-propelled
particles at $t$, and it shall be }as similar as possible to the
target $\mathbf{S}(t+\Delta t)$, i.e., the position and velocity
configuration obtained directly by the simulations. Here, we fix $\eta=0.2,L=5,R=1$
for instance, \textcolor{black}{and assume periodic boundary conditions.
The system is initialized in the disordered phase, and such random
states }are provided to train the NANN, with any $t>0$ being a subsequent
time step to be predicted.

Fig.~\ref{Vicsek}(a) shows a direct comparison of the time evolution
of the order parameter $v_{a}$. The black curve is the $v_{a}$ calculated
using the results of the real simulations on the $N=100$ system,
and the red curve is the $v_{a}$ calculated using NANN's outcomes
for the same system size. The \textcolor{black}{consistency in }$v_{a}$
manifests that, given only the $t=0$ one-step data for training,
NANN has the effectiveness to predict this nonequilibrium complex
system's long-term dynamics. In particular, the system's long-term
flocking dynamics appear quite differently and more complicatedly
compared with the one-step evolution at $t=0$, but NANN's effectiveness
remains robust to a considerable extent. This can be expected only
when the machine learning model does learn the underlying logic of
Eq.~(\ref{eq:Direction_with_extrinsic_noise}), rather than merely
learning specific patterns, configurations, or evolutionary trajectories.\textcolor{black}{{}
}Furthermore, from Fig.~\ref{Vicsek}(b) one can clearly see the
extrapolation capability of NANN also remaining robust for this nonequilibrium
complex system. NANN \textcolor{black}{successfully} generates the
long-term dynamics of the $N_{+}=200$ system by just training on
$N=100$. \textcolor{black}{The }time evolution of\textcolor{black}{{}
the group velocity }$v_{a}$ of all $N$ self-propelled particles
\textcolor{black}{during the} flocking dynamics\textcolor{black}{{}
reveal no considerable difference between NANN's }extrapolating outputs
and the results of the real simulations on Eq.~(\ref{eq:Direction_with_extrinsic_noise}).
\textcolor{black}{These machine learning results for both the Vicsek
model and the noisy Kuramoto model are obtained }in a quite generic
manner without any case-by-case special design on the network architecture,
strongly suggesting that NANN can be readily applied for various complex
systems.

In general, we find that the efficiency of using NANN to generate
evolution trajectories can be about one order of magnitude higher
than that of using the original algorithm to simulate the stochastic
dynamics simulations on the same system size. On this foundation,
when specifically applying NANN to any certain physical system, its
efficiency can be further improved by providing more physical information
to the machine learning model. Consider the noisy Kuramoto model for
instance. Noticing that the interactions are global in that case,
the attention weight $W$ in the architecture (see also Fig.~\ref{NANN})
that models the strength of interactions among all particles can be
set to unity by hand, which significantly improves the efficiency.
Finally, it is worth mentioning that there of course exist some conventional
but skillfully optimized algorithms that are even faster. It is common
knowledge that artificial intelligence is currently unable to really
master or even replace conventional human wisdom in physics. Like
many machine learning applications in physics \citep{Farimani_JCP_2022,behler_PRL_2007,Chmiela_SciAdv_2017,Zhanglinfeng_PRL_2018,Zhanglinfeng_MP_2019,Zhanglinfeng_JCP_2021,Zhanglinfeng_CPC_2021,Zhongzhicheng_PRB_2022,Wanglei_PRX_2020,Wanglei_JML_2022,Wanglei_SciPost_2023,Carleo_RMP_2019,Carrasquilla_AdvPhysX_2020,Volpe_Nat_Mach_Intell_2020,Melko_Nat_Phys_2017,van_Nieuwenburg_Nat_Phys_2017,Wanglei_PRL_2018_RG,Guo_NJP_2023},
NANN is a cooperator of those conventional ways, rather than a competitor.
Taking the advantages of machine learning in terms of universality
and transferability, NANN is intended to serve as a generic tool and
help physicists more conveniently find the interesting physical systems
and key parameter regions that are of special value and significance
to be focused on and investigated in the conventional ways. In this
practical sense, we believe that NANN can stimulate various investigations
in all fields associated with stochastic dynamics simulations.

\section{Conclusions }

We propose a new type of neural network, noise-aware neural network
(NANN), based on which a generic machine learning approach is developed
to simulate the stochastic dynamics, as demonstrated in the noisy
Kuramoto model and the Vicsek model. A key feature of this approach
is its ability to generate the long-term stochastic dynamics of complex
large-scale systems by training NANN with the one-step dynamics of
the smaller-scale systems. This can be particularly useful in investigating
long-time dynamical behavior of large complex systems in different
fields, ranging from active matter \citep{Solon_PRL_2015,Omar_PRL_2021,wysocki_EuroLett_2014,siebert_Soft_2017,Digregorio_PRL_2018,Speck_PRE_2021,fernandez_NatCom_2020,rein_EuroLett_2023,Vicsek_PRL_1995,Chate_PRL_2004,Toner_Ann_Phys_2005,Chate_PRE_2008},
over intelligent matter \citep{Boguna_PRE_2015,Boguna_PRE_2017,Yusufaly_PRE_2016_QuorumSensing,Aguilar_PRE_2021_QuorumSensing,Tobias_Nat_Commun_2018_QuorumSensing,Lowen_PNAS_2021},
to social physics \citep{farkas_physA_2003} and etc., where rich
phenomena, such as the cross sea pattern of self-propelled particles
\citep{kursten_PRL_2020}, the band pattern of self-propelled rods
\citep{ginelli_PRL_2010}, and the large-scale vortex of moving microtubules
\citep{sumino_Nat_2012}, etc., could emerge. 
\begin{acknowledgments}
This work is supported by NKRDPC (Grant No.~2022YFA1405304), NSFC
(Grant No.~12275089), Guangdong Basic and Applied Basic Research
Foundation (Grants No.~2023A1515012800), Guangdong Provincial Key
Laboratory (Grant No.~2020B1212060066).
\end{acknowledgments}

\bibliographystyle{apsrev4-1}

\begin{thebibliography}{56}%
\makeatletter
\providecommand \@ifxundefined [1]{%
 \@ifx{#1\undefined}
}%
\providecommand \@ifnum [1]{%
 \ifnum #1\expandafter \@firstoftwo
 \else \expandafter \@secondoftwo
 \fi
}%
\providecommand \@ifx [1]{%
 \ifx #1\expandafter \@firstoftwo
 \else \expandafter \@secondoftwo
 \fi
}%
\providecommand \natexlab [1]{#1}%
\providecommand \enquote  [1]{``#1''}%
\providecommand \bibnamefont  [1]{#1}%
\providecommand \bibfnamefont [1]{#1}%
\providecommand \citenamefont [1]{#1}%
\providecommand \href@noop [0]{\@secondoftwo}%
\providecommand \href [0]{\begingroup \@sanitize@url \@href}%
\providecommand \@href[1]{\@@startlink{#1}\@@href}%
\providecommand \@@href[1]{\endgroup#1\@@endlink}%
\providecommand \@sanitize@url [0]{\catcode `\\12\catcode `\$12\catcode
  `\&12\catcode `\#12\catcode `\^12\catcode `\_12\catcode `\%12\relax}%
\providecommand \@@startlink[1]{}%
\providecommand \@@endlink[0]{}%
\providecommand \url  [0]{\begingroup\@sanitize@url \@url }%
\providecommand \@url [1]{\endgroup\@href {#1}{\urlprefix }}%
\providecommand \urlprefix  [0]{URL }%
\providecommand \Eprint [0]{\href }%
\providecommand \doibase [0]{http://dx.doi.org/}%
\providecommand \selectlanguage [0]{\@gobble}%
\providecommand \bibinfo  [0]{\@secondoftwo}%
\providecommand \bibfield  [0]{\@secondoftwo}%
\providecommand \translation [1]{[#1]}%
\providecommand \BibitemOpen [0]{}%
\providecommand \bibitemStop [0]{}%
\providecommand \bibitemNoStop [0]{.\EOS\space}%
\providecommand \EOS [0]{\spacefactor3000\relax}%
\providecommand \BibitemShut  [1]{\csname bibitem#1\endcsname}%
\let\auto@bib@innerbib\@empty
\bibitem [{\citenamefont {Li}\ \emph {et~al.}(2022)\citenamefont {Li},
  \citenamefont {Meidani}, \citenamefont {Yadav},\ and\ \citenamefont
  {Barati~Farimani}}]{Farimani_JCP_2022}%
  \BibitemOpen
  \bibfield  {author} {\bibinfo {author} {\bibfnamefont {Z.}~\bibnamefont
  {Li}}, \bibinfo {author} {\bibfnamefont {K.}~\bibnamefont {Meidani}},
  \bibinfo {author} {\bibfnamefont {P.}~\bibnamefont {Yadav}}, \ and\ \bibinfo
  {author} {\bibfnamefont {A.}~\bibnamefont {Barati~Farimani}},\ }\href
  {\doibase 10.1063/5.0083060} {\bibfield  {journal} {\bibinfo  {journal} {J.
  Chem. Phys.}\ }\textbf {\bibinfo {volume} {156}},\ \bibinfo {pages} {144103}
  (\bibinfo {year} {2022})}\BibitemShut {NoStop}%
\bibitem [{\citenamefont {Behler}\ and\ \citenamefont
  {Parrinello}(2007)}]{behler_PRL_2007}%
  \BibitemOpen
  \bibfield  {author} {\bibinfo {author} {\bibfnamefont {J.}~\bibnamefont
  {Behler}}\ and\ \bibinfo {author} {\bibfnamefont {M.}~\bibnamefont
  {Parrinello}},\ }\href {\doibase 10.1103/PhysRevLett.98.146401} {\bibfield
  {journal} {\bibinfo  {journal} {Phys. Rev. Lett.}\ }\textbf {\bibinfo
  {volume} {98}},\ \bibinfo {pages} {146401} (\bibinfo {year}
  {2007})}\BibitemShut {NoStop}%
\bibitem [{\citenamefont {Chmiela}\ \emph {et~al.}(2017)\citenamefont
  {Chmiela}, \citenamefont {Tkatchenko}, \citenamefont {Sauceda}, \citenamefont
  {Poltavsky}, \citenamefont {Sch{\"u}tt},\ and\ \citenamefont
  {M{\"u}ller}}]{Chmiela_SciAdv_2017}%
  \BibitemOpen
  \bibfield  {author} {\bibinfo {author} {\bibfnamefont {S.}~\bibnamefont
  {Chmiela}}, \bibinfo {author} {\bibfnamefont {A.}~\bibnamefont {Tkatchenko}},
  \bibinfo {author} {\bibfnamefont {H.~E.}\ \bibnamefont {Sauceda}}, \bibinfo
  {author} {\bibfnamefont {I.}~\bibnamefont {Poltavsky}}, \bibinfo {author}
  {\bibfnamefont {K.~T.}\ \bibnamefont {Sch{\"u}tt}}, \ and\ \bibinfo {author}
  {\bibfnamefont {K.-R.}\ \bibnamefont {M{\"u}ller}},\ }\href {\doibase
  10.1126/sciadv.1603015} {\bibfield  {journal} {\bibinfo  {journal} {Sci.
  Adv.}\ }\textbf {\bibinfo {volume} {3}},\ \bibinfo {pages} {e1603015}
  (\bibinfo {year} {2017})}\BibitemShut {NoStop}%
\bibitem [{\citenamefont {Zhang}\ \emph {et~al.}(2018)\citenamefont {Zhang},
  \citenamefont {Han}, \citenamefont {Wang}, \citenamefont {Car},\ and\
  \citenamefont {E}}]{Zhanglinfeng_PRL_2018}%
  \BibitemOpen
  \bibfield  {author} {\bibinfo {author} {\bibfnamefont {L.}~\bibnamefont
  {Zhang}}, \bibinfo {author} {\bibfnamefont {J.}~\bibnamefont {Han}}, \bibinfo
  {author} {\bibfnamefont {H.}~\bibnamefont {Wang}}, \bibinfo {author}
  {\bibfnamefont {R.}~\bibnamefont {Car}}, \ and\ \bibinfo {author}
  {\bibfnamefont {W.}~\bibnamefont {E}},\ }\href {\doibase
  10.1103/PhysRevLett.120.143001} {\bibfield  {journal} {\bibinfo  {journal}
  {Phys. Rev. Lett.}\ }\textbf {\bibinfo {volume} {120}},\ \bibinfo {pages}
  {143001} (\bibinfo {year} {2018})}\BibitemShut {NoStop}%
\bibitem [{\citenamefont {Ko}\ \emph {et~al.}(2019)\citenamefont {Ko},
  \citenamefont {Zhang}, \citenamefont {Santra}, \citenamefont {Wang},
  \citenamefont {E}, \citenamefont {DiStasio},\ and\ \citenamefont
  {Car}}]{Zhanglinfeng_MP_2019}%
  \BibitemOpen
  \bibfield  {author} {\bibinfo {author} {\bibfnamefont {H.-Y.}\ \bibnamefont
  {Ko}}, \bibinfo {author} {\bibfnamefont {L.}~\bibnamefont {Zhang}}, \bibinfo
  {author} {\bibfnamefont {B.}~\bibnamefont {Santra}}, \bibinfo {author}
  {\bibfnamefont {H.}~\bibnamefont {Wang}}, \bibinfo {author} {\bibfnamefont
  {W.}~\bibnamefont {E}}, \bibinfo {author} {\bibfnamefont {R.~A.~J.}\
  \bibnamefont {DiStasio}}, \ and\ \bibinfo {author} {\bibfnamefont
  {R.}~\bibnamefont {Car}},\ }\href {\doibase 10.1080/00268976.2019.1652366}
  {\bibfield  {journal} {\bibinfo  {journal} {Mol. Phys.}\ }\textbf {\bibinfo
  {volume} {117}},\ \bibinfo {pages} {3269} (\bibinfo {year}
  {2019})}\BibitemShut {NoStop}%
\bibitem [{\citenamefont {Huang}\ \emph {et~al.}(2021)\citenamefont {Huang},
  \citenamefont {Zhang}, \citenamefont {Wang}, \citenamefont {Zhao},
  \citenamefont {Cheng},\ and\ \citenamefont {E}}]{Zhanglinfeng_JCP_2021}%
  \BibitemOpen
  \bibfield  {author} {\bibinfo {author} {\bibfnamefont {J.}~\bibnamefont
  {Huang}}, \bibinfo {author} {\bibfnamefont {L.}~\bibnamefont {Zhang}},
  \bibinfo {author} {\bibfnamefont {H.}~\bibnamefont {Wang}}, \bibinfo {author}
  {\bibfnamefont {J.}~\bibnamefont {Zhao}}, \bibinfo {author} {\bibfnamefont
  {J.}~\bibnamefont {Cheng}}, \ and\ \bibinfo {author} {\bibfnamefont
  {W.}~\bibnamefont {E}},\ }\href {\doibase 10.1063/5.0041849} {\bibfield
  {journal} {\bibinfo  {journal} {J. Chem. Phys.}\ }\textbf {\bibinfo {volume}
  {154}},\ \bibinfo {pages} {094703} (\bibinfo {year} {2021})}\BibitemShut
  {NoStop}%
\bibitem [{\citenamefont {Lu}\ \emph {et~al.}(2021)\citenamefont {Lu},
  \citenamefont {Wang}, \citenamefont {Chen}, \citenamefont {Lin},
  \citenamefont {Car}, \citenamefont {E}, \citenamefont {Jia},\ and\
  \citenamefont {Zhang}}]{Zhanglinfeng_CPC_2021}%
  \BibitemOpen
  \bibfield  {author} {\bibinfo {author} {\bibfnamefont {D.}~\bibnamefont
  {Lu}}, \bibinfo {author} {\bibfnamefont {H.}~\bibnamefont {Wang}}, \bibinfo
  {author} {\bibfnamefont {M.}~\bibnamefont {Chen}}, \bibinfo {author}
  {\bibfnamefont {L.}~\bibnamefont {Lin}}, \bibinfo {author} {\bibfnamefont
  {R.}~\bibnamefont {Car}}, \bibinfo {author} {\bibfnamefont {W.}~\bibnamefont
  {E}}, \bibinfo {author} {\bibfnamefont {W.}~\bibnamefont {Jia}}, \ and\
  \bibinfo {author} {\bibfnamefont {L.}~\bibnamefont {Zhang}},\ }\href
  {\doibase https://doi.org/10.1016/j.cpc.2020.107624} {\bibfield  {journal}
  {\bibinfo  {journal} {Comput. Phys. Commun.}\ }\textbf {\bibinfo {volume}
  {259}},\ \bibinfo {pages} {107624} (\bibinfo {year} {2021})}\BibitemShut
  {NoStop}%
\bibitem [{\citenamefont {He}\ \emph {et~al.}(2022)\citenamefont {He},
  \citenamefont {Wu}, \citenamefont {Zhang}, \citenamefont {Wang},
  \citenamefont {Fu}, \citenamefont {Liu},\ and\ \citenamefont
  {Zhong}}]{Zhongzhicheng_PRB_2022}%
  \BibitemOpen
  \bibfield  {author} {\bibinfo {author} {\bibfnamefont {R.}~\bibnamefont
  {He}}, \bibinfo {author} {\bibfnamefont {H.}~\bibnamefont {Wu}}, \bibinfo
  {author} {\bibfnamefont {L.}~\bibnamefont {Zhang}}, \bibinfo {author}
  {\bibfnamefont {X.}~\bibnamefont {Wang}}, \bibinfo {author} {\bibfnamefont
  {F.}~\bibnamefont {Fu}}, \bibinfo {author} {\bibfnamefont {S.}~\bibnamefont
  {Liu}}, \ and\ \bibinfo {author} {\bibfnamefont {Z.}~\bibnamefont {Zhong}},\
  }\href {\doibase 10.1103/PhysRevB.105.064104} {\bibfield  {journal} {\bibinfo
   {journal} {Phys. Rev. B}\ }\textbf {\bibinfo {volume} {105}},\ \bibinfo
  {pages} {064104} (\bibinfo {year} {2022})}\BibitemShut {NoStop}%
\bibitem [{\citenamefont {Li}\ \emph {et~al.}(2020)\citenamefont {Li},
  \citenamefont {Dong}, \citenamefont {Zhang},\ and\ \citenamefont
  {Wang}}]{Wanglei_PRX_2020}%
  \BibitemOpen
  \bibfield  {author} {\bibinfo {author} {\bibfnamefont {S.-H.}\ \bibnamefont
  {Li}}, \bibinfo {author} {\bibfnamefont {C.-X.}\ \bibnamefont {Dong}},
  \bibinfo {author} {\bibfnamefont {L.}~\bibnamefont {Zhang}}, \ and\ \bibinfo
  {author} {\bibfnamefont {L.}~\bibnamefont {Wang}},\ }\href {\doibase
  10.1103/PhysRevX.10.021020} {\bibfield  {journal} {\bibinfo  {journal} {Phys.
  Rev. X}\ }\textbf {\bibinfo {volume} {10}},\ \bibinfo {pages} {021020}
  (\bibinfo {year} {2020})}\BibitemShut {NoStop}%
\bibitem [{\citenamefont {Xie}\ \emph {et~al.}(2022)\citenamefont {Xie},
  \citenamefont {Zhang},\ and\ \citenamefont {Wang}}]{Wanglei_JML_2022}%
  \BibitemOpen
  \bibfield  {author} {\bibinfo {author} {\bibfnamefont {H.}~\bibnamefont
  {Xie}}, \bibinfo {author} {\bibfnamefont {L.}~\bibnamefont {Zhang}}, \ and\
  \bibinfo {author} {\bibfnamefont {L.}~\bibnamefont {Wang}},\ }\href {\doibase
  https://doi.org/10.4208/jml.220113} {\bibfield  {journal} {\bibinfo
  {journal} {J. Mach. Learn.}\ }\textbf {\bibinfo {volume} {1}},\ \bibinfo
  {pages} {38} (\bibinfo {year} {2022})}\BibitemShut {NoStop}%
\bibitem [{\citenamefont {Xie}\ \emph {et~al.}(2023)\citenamefont {Xie},
  \citenamefont {Zhang},\ and\ \citenamefont {Wang}}]{Wanglei_SciPost_2023}%
  \BibitemOpen
  \bibfield  {author} {\bibinfo {author} {\bibfnamefont {H.}~\bibnamefont
  {Xie}}, \bibinfo {author} {\bibfnamefont {L.}~\bibnamefont {Zhang}}, \ and\
  \bibinfo {author} {\bibfnamefont {L.}~\bibnamefont {Wang}},\ }\href {\doibase
  10.21468/SciPostPhys.14.6.154} {\bibfield  {journal} {\bibinfo  {journal}
  {SciPost Phys.}\ }\textbf {\bibinfo {volume} {14}},\ \bibinfo {pages} {154}
  (\bibinfo {year} {2023})}\BibitemShut {NoStop}%
\bibitem [{\citenamefont {Honerkamp}(1996)}]{honerkamp_book_1996}%
  \BibitemOpen
  \bibfield  {author} {\bibinfo {author} {\bibfnamefont {J.}~\bibnamefont
  {Honerkamp}},\ }\href@noop {} {\emph {\bibinfo {title} {Stochastic dynamical
  systems: concepts, numerical methods, data analysis}}}\ (\bibinfo
  {publisher} {John Wiley \& Sons},\ \bibinfo {year} {1996})\BibitemShut
  {NoStop}%
\bibitem [{\citenamefont {Casert}\ \emph {et~al.}()\citenamefont {Casert},
  \citenamefont {Tamblyn},\ and\ \citenamefont {Whitelam}}]{Casert_arXiv_2022}%
  \BibitemOpen
  \bibfield  {author} {\bibinfo {author} {\bibfnamefont {C.}~\bibnamefont
  {Casert}}, \bibinfo {author} {\bibfnamefont {I.}~\bibnamefont {Tamblyn}}, \
  and\ \bibinfo {author} {\bibfnamefont {S.}~\bibnamefont {Whitelam}},\
  }\href@noop {} {\ }\Eprint {http://arxiv.org/abs/2202.08708}
  {arXiv:2202.08708} \BibitemShut {NoStop}%
\bibitem [{\citenamefont {Schmitt}\ \emph {et~al.}()\citenamefont {Schmitt},
  \citenamefont {Koch-Janusz}, \citenamefont {Fruchart}, \citenamefont
  {Seara},\ and\ \citenamefont {Vitelli}}]{MKJ_arXiv_2023}%
  \BibitemOpen
  \bibfield  {author} {\bibinfo {author} {\bibfnamefont {M.~S.}\ \bibnamefont
  {Schmitt}}, \bibinfo {author} {\bibfnamefont {M.}~\bibnamefont
  {Koch-Janusz}}, \bibinfo {author} {\bibfnamefont {M.}~\bibnamefont
  {Fruchart}}, \bibinfo {author} {\bibfnamefont {D.~S.}\ \bibnamefont {Seara}},
  \ and\ \bibinfo {author} {\bibfnamefont {V.}~\bibnamefont {Vitelli}},\
  }\href@noop {} {\ }\Eprint {http://arxiv.org/abs/2312.06608}
  {arXiv:2312.06608} \BibitemShut {NoStop}%
\bibitem [{\citenamefont {Ha}\ and\ \citenamefont
  {Jeong}(2021)}]{Jeong_SciRep_2021}%
  \BibitemOpen
  \bibfield  {author} {\bibinfo {author} {\bibfnamefont {S.}~\bibnamefont
  {Ha}}\ and\ \bibinfo {author} {\bibfnamefont {H.}~\bibnamefont {Jeong}},\
  }\href {\doibase 10.1038/s41598-021-91878-w} {\bibfield  {journal} {\bibinfo
  {journal} {Sci. Rep.}\ }\textbf {\bibinfo {volume} {11}},\ \bibinfo {pages}
  {12804} (\bibinfo {year} {2021})}\BibitemShut {NoStop}%
\bibitem [{\citenamefont {Zhu}\ \emph {et~al.}(2023)\citenamefont {Zhu},
  \citenamefont {Tang},\ and\ \citenamefont {Kim}}]{Zhu_JCP_2023}%
  \BibitemOpen
  \bibfield  {author} {\bibinfo {author} {\bibfnamefont {Y.}~\bibnamefont
  {Zhu}}, \bibinfo {author} {\bibfnamefont {Y.-H.}\ \bibnamefont {Tang}}, \
  and\ \bibinfo {author} {\bibfnamefont {C.}~\bibnamefont {Kim}},\ }\href
  {\doibase https://doi.org/10.1016/j.jcp.2022.111819} {\bibfield  {journal}
  {\bibinfo  {journal} {J. Comput. Phys.}\ }\textbf {\bibinfo {volume} {474}},\
  \bibinfo {pages} {111819} (\bibinfo {year} {2023})}\BibitemShut {NoStop}%
\bibitem [{\citenamefont {Solon}\ \emph {et~al.}(2015)\citenamefont {Solon},
  \citenamefont {Stenhammar}, \citenamefont {Wittkowski}, \citenamefont
  {Kardar}, \citenamefont {Kafri}, \citenamefont {Cates},\ and\ \citenamefont
  {Tailleur}}]{Solon_PRL_2015}%
  \BibitemOpen
  \bibfield  {author} {\bibinfo {author} {\bibfnamefont {A.~P.}\ \bibnamefont
  {Solon}}, \bibinfo {author} {\bibfnamefont {J.}~\bibnamefont {Stenhammar}},
  \bibinfo {author} {\bibfnamefont {R.}~\bibnamefont {Wittkowski}}, \bibinfo
  {author} {\bibfnamefont {M.}~\bibnamefont {Kardar}}, \bibinfo {author}
  {\bibfnamefont {Y.}~\bibnamefont {Kafri}}, \bibinfo {author} {\bibfnamefont
  {M.~E.}\ \bibnamefont {Cates}}, \ and\ \bibinfo {author} {\bibfnamefont
  {J.}~\bibnamefont {Tailleur}},\ }\href {\doibase
  10.1103/PhysRevLett.114.198301} {\bibfield  {journal} {\bibinfo  {journal}
  {Phys. Rev. Lett.}\ }\textbf {\bibinfo {volume} {114}},\ \bibinfo {pages}
  {198301} (\bibinfo {year} {2015})}\BibitemShut {NoStop}%
\bibitem [{\citenamefont {Omar}\ \emph {et~al.}(2021)\citenamefont {Omar},
  \citenamefont {Klymko}, \citenamefont {GrandPre},\ and\ \citenamefont
  {Geissler}}]{Omar_PRL_2021}%
  \BibitemOpen
  \bibfield  {author} {\bibinfo {author} {\bibfnamefont {A.~K.}\ \bibnamefont
  {Omar}}, \bibinfo {author} {\bibfnamefont {K.}~\bibnamefont {Klymko}},
  \bibinfo {author} {\bibfnamefont {T.}~\bibnamefont {GrandPre}}, \ and\
  \bibinfo {author} {\bibfnamefont {P.~L.}\ \bibnamefont {Geissler}},\ }\href
  {\doibase 10.1103/PhysRevLett.126.188002} {\bibfield  {journal} {\bibinfo
  {journal} {Phys. Rev. Lett.}\ }\textbf {\bibinfo {volume} {126}},\ \bibinfo
  {pages} {188002} (\bibinfo {year} {2021})}\BibitemShut {NoStop}%
\bibitem [{\citenamefont {Wysocki}\ \emph {et~al.}(2014)\citenamefont
  {Wysocki}, \citenamefont {Winkler},\ and\ \citenamefont
  {Gompper}}]{wysocki_EuroLett_2014}%
  \BibitemOpen
  \bibfield  {author} {\bibinfo {author} {\bibfnamefont {A.}~\bibnamefont
  {Wysocki}}, \bibinfo {author} {\bibfnamefont {R.~G.}\ \bibnamefont
  {Winkler}}, \ and\ \bibinfo {author} {\bibfnamefont {G.}~\bibnamefont
  {Gompper}},\ }\href {\doibase 10.1209/0295-5075/105/48004} {\bibfield
  {journal} {\bibinfo  {journal} {EPL}\ }\textbf {\bibinfo {volume} {105}},\
  \bibinfo {pages} {48004} (\bibinfo {year} {2014})}\BibitemShut {NoStop}%
\bibitem [{\citenamefont {Siebert}\ \emph {et~al.}(2017)\citenamefont
  {Siebert}, \citenamefont {Letz}, \citenamefont {Speck},\ and\ \citenamefont
  {Virnau}}]{siebert_Soft_2017}%
  \BibitemOpen
  \bibfield  {author} {\bibinfo {author} {\bibfnamefont {J.~T.}\ \bibnamefont
  {Siebert}}, \bibinfo {author} {\bibfnamefont {J.}~\bibnamefont {Letz}},
  \bibinfo {author} {\bibfnamefont {T.}~\bibnamefont {Speck}}, \ and\ \bibinfo
  {author} {\bibfnamefont {P.}~\bibnamefont {Virnau}},\ }\href {\doibase
  10.1039/C6SM02622B} {\bibfield  {journal} {\bibinfo  {journal} {Soft Matter}\
  }\textbf {\bibinfo {volume} {13}},\ \bibinfo {pages} {1020} (\bibinfo {year}
  {2017})}\BibitemShut {NoStop}%
\bibitem [{\citenamefont {Digregorio}\ \emph {et~al.}(2018)\citenamefont
  {Digregorio}, \citenamefont {Levis}, \citenamefont {Suma}, \citenamefont
  {Cugliandolo}, \citenamefont {Gonnella},\ and\ \citenamefont
  {Pagonabarraga}}]{Digregorio_PRL_2018}%
  \BibitemOpen
  \bibfield  {author} {\bibinfo {author} {\bibfnamefont {P.}~\bibnamefont
  {Digregorio}}, \bibinfo {author} {\bibfnamefont {D.}~\bibnamefont {Levis}},
  \bibinfo {author} {\bibfnamefont {A.}~\bibnamefont {Suma}}, \bibinfo {author}
  {\bibfnamefont {L.~F.}\ \bibnamefont {Cugliandolo}}, \bibinfo {author}
  {\bibfnamefont {G.}~\bibnamefont {Gonnella}}, \ and\ \bibinfo {author}
  {\bibfnamefont {I.}~\bibnamefont {Pagonabarraga}},\ }\href {\doibase
  10.1103/PhysRevLett.121.098003} {\bibfield  {journal} {\bibinfo  {journal}
  {Phys. Rev. Lett.}\ }\textbf {\bibinfo {volume} {121}},\ \bibinfo {pages}
  {098003} (\bibinfo {year} {2018})}\BibitemShut {NoStop}%
\bibitem [{\citenamefont {Speck}(2021)}]{Speck_PRE_2021}%
  \BibitemOpen
  \bibfield  {author} {\bibinfo {author} {\bibfnamefont {T.}~\bibnamefont
  {Speck}},\ }\href {\doibase 10.1103/PhysRevE.103.012607} {\bibfield
  {journal} {\bibinfo  {journal} {Phys. Rev. E}\ }\textbf {\bibinfo {volume}
  {103}},\ \bibinfo {pages} {012607} (\bibinfo {year} {2021})}\BibitemShut
  {NoStop}%
\bibitem [{\citenamefont {Fernandez-Rodriguez}\ \emph
  {et~al.}(2020)\citenamefont {Fernandez-Rodriguez}, \citenamefont {Grillo},
  \citenamefont {Alvarez}, \citenamefont {Rathlef}, \citenamefont {Buttinoni},
  \citenamefont {Volpe},\ and\ \citenamefont {Isa}}]{fernandez_NatCom_2020}%
  \BibitemOpen
  \bibfield  {author} {\bibinfo {author} {\bibfnamefont {M.~A.}\ \bibnamefont
  {Fernandez-Rodriguez}}, \bibinfo {author} {\bibfnamefont {F.}~\bibnamefont
  {Grillo}}, \bibinfo {author} {\bibfnamefont {L.}~\bibnamefont {Alvarez}},
  \bibinfo {author} {\bibfnamefont {M.}~\bibnamefont {Rathlef}}, \bibinfo
  {author} {\bibfnamefont {I.}~\bibnamefont {Buttinoni}}, \bibinfo {author}
  {\bibfnamefont {G.}~\bibnamefont {Volpe}}, \ and\ \bibinfo {author}
  {\bibfnamefont {L.}~\bibnamefont {Isa}},\ }\href
  {https://www.nature.com/articles/s41467-020-17864-4} {\bibfield  {journal}
  {\bibinfo  {journal} {Nat. Commun.}\ }\textbf {\bibinfo {volume} {11}},\
  \bibinfo {pages} {4223} (\bibinfo {year} {2020})}\BibitemShut {NoStop}%
\bibitem [{\citenamefont {Rein}\ \emph {et~al.}(2023)\citenamefont {Rein},
  \citenamefont {Kol{\'a}{\v{r}}}, \citenamefont {Kroy},\ and\ \citenamefont
  {Holubec}}]{rein_EuroLett_2023}%
  \BibitemOpen
  \bibfield  {author} {\bibinfo {author} {\bibfnamefont {C.}~\bibnamefont
  {Rein}}, \bibinfo {author} {\bibfnamefont {M.}~\bibnamefont
  {Kol{\'a}{\v{r}}}}, \bibinfo {author} {\bibfnamefont {K.}~\bibnamefont
  {Kroy}}, \ and\ \bibinfo {author} {\bibfnamefont {V.}~\bibnamefont
  {Holubec}},\ }\href {\doibase 10.1209/0295-5075/accca5} {\bibfield  {journal}
  {\bibinfo  {journal} {EPL}\ }\textbf {\bibinfo {volume} {142}},\ \bibinfo
  {pages} {31001} (\bibinfo {year} {2023})}\BibitemShut {NoStop}%
\bibitem [{\citenamefont {Vicsek}\ \emph {et~al.}(1995)\citenamefont {Vicsek},
  \citenamefont {Czir\'ok}, \citenamefont {Ben-Jacob}, \citenamefont {Cohen},\
  and\ \citenamefont {Shochet}}]{Vicsek_PRL_1995}%
  \BibitemOpen
  \bibfield  {author} {\bibinfo {author} {\bibfnamefont {T.}~\bibnamefont
  {Vicsek}}, \bibinfo {author} {\bibfnamefont {A.}~\bibnamefont {Czir\'ok}},
  \bibinfo {author} {\bibfnamefont {E.}~\bibnamefont {Ben-Jacob}}, \bibinfo
  {author} {\bibfnamefont {I.}~\bibnamefont {Cohen}}, \ and\ \bibinfo {author}
  {\bibfnamefont {O.}~\bibnamefont {Shochet}},\ }\href {\doibase
  10.1103/PhysRevLett.75.1226} {\bibfield  {journal} {\bibinfo  {journal}
  {Phys. Rev. Lett.}\ }\textbf {\bibinfo {volume} {75}},\ \bibinfo {pages}
  {1226} (\bibinfo {year} {1995})}\BibitemShut {NoStop}%
\bibitem [{\citenamefont {Gr\'egoire}\ and\ \citenamefont
  {Chat\'e}(2004)}]{Chate_PRL_2004}%
  \BibitemOpen
  \bibfield  {author} {\bibinfo {author} {\bibfnamefont {G.}~\bibnamefont
  {Gr\'egoire}}\ and\ \bibinfo {author} {\bibfnamefont {H.}~\bibnamefont
  {Chat\'e}},\ }\href {\doibase 10.1103/PhysRevLett.92.025702} {\bibfield
  {journal} {\bibinfo  {journal} {Phys. Rev. Lett.}\ }\textbf {\bibinfo
  {volume} {92}},\ \bibinfo {pages} {025702} (\bibinfo {year}
  {2004})}\BibitemShut {NoStop}%
\bibitem [{\citenamefont {Toner}\ \emph {et~al.}(2005)\citenamefont {Toner},
  \citenamefont {Tu},\ and\ \citenamefont {Ramaswamy}}]{Toner_Ann_Phys_2005}%
  \BibitemOpen
  \bibfield  {author} {\bibinfo {author} {\bibfnamefont {J.}~\bibnamefont
  {Toner}}, \bibinfo {author} {\bibfnamefont {Y.}~\bibnamefont {Tu}}, \ and\
  \bibinfo {author} {\bibfnamefont {S.}~\bibnamefont {Ramaswamy}},\ }\href
  {\doibase https://doi.org/10.1016/j.aop.2005.04.011} {\bibfield  {journal}
  {\bibinfo  {journal} {Ann. Phys.}\ }\textbf {\bibinfo {volume} {318}},\
  \bibinfo {pages} {170} (\bibinfo {year} {2005})}\BibitemShut {NoStop}%
\bibitem [{\citenamefont {Chat\'e}\ \emph {et~al.}(2008)\citenamefont
  {Chat\'e}, \citenamefont {Ginelli}, \citenamefont {Gr\'egoire},\ and\
  \citenamefont {Raynaud}}]{Chate_PRE_2008}%
  \BibitemOpen
  \bibfield  {author} {\bibinfo {author} {\bibfnamefont {H.}~\bibnamefont
  {Chat\'e}}, \bibinfo {author} {\bibfnamefont {F.}~\bibnamefont {Ginelli}},
  \bibinfo {author} {\bibfnamefont {G.}~\bibnamefont {Gr\'egoire}}, \ and\
  \bibinfo {author} {\bibfnamefont {F.}~\bibnamefont {Raynaud}},\ }\href
  {\doibase 10.1103/PhysRevE.77.046113} {\bibfield  {journal} {\bibinfo
  {journal} {Phys. Rev. E}\ }\textbf {\bibinfo {volume} {77}},\ \bibinfo
  {pages} {046113} (\bibinfo {year} {2008})}\BibitemShut {NoStop}%
\bibitem [{\citenamefont {Mosquera-Do\~nate}\ and\ \citenamefont
  {Bogu\~n\'a}(2015)}]{Boguna_PRE_2015}%
  \BibitemOpen
  \bibfield  {author} {\bibinfo {author} {\bibfnamefont {G.}~\bibnamefont
  {Mosquera-Do\~nate}}\ and\ \bibinfo {author} {\bibfnamefont {M.}~\bibnamefont
  {Bogu\~n\'a}},\ }\href {\doibase 10.1103/PhysRevE.91.052804} {\bibfield
  {journal} {\bibinfo  {journal} {Phys. Rev. E}\ }\textbf {\bibinfo {volume}
  {91}},\ \bibinfo {pages} {052804} (\bibinfo {year} {2015})}\BibitemShut
  {NoStop}%
\bibitem [{\citenamefont {Rozanova}\ and\ \citenamefont
  {Bogu\~n\'a}(2017)}]{Boguna_PRE_2017}%
  \BibitemOpen
  \bibfield  {author} {\bibinfo {author} {\bibfnamefont {L.}~\bibnamefont
  {Rozanova}}\ and\ \bibinfo {author} {\bibfnamefont {M.}~\bibnamefont
  {Bogu\~n\'a}},\ }\href {\doibase 10.1103/PhysRevE.96.012310} {\bibfield
  {journal} {\bibinfo  {journal} {Phys. Rev. E}\ }\textbf {\bibinfo {volume}
  {96}},\ \bibinfo {pages} {012310} (\bibinfo {year} {2017})}\BibitemShut
  {NoStop}%
\bibitem [{\citenamefont {Yusufaly}\ and\ \citenamefont
  {Boedicker}(2016)}]{Yusufaly_PRE_2016_QuorumSensing}%
  \BibitemOpen
  \bibfield  {author} {\bibinfo {author} {\bibfnamefont {T.~I.}\ \bibnamefont
  {Yusufaly}}\ and\ \bibinfo {author} {\bibfnamefont {J.~Q.}\ \bibnamefont
  {Boedicker}},\ }\href {\doibase 10.1103/PhysRevE.94.062410} {\bibfield
  {journal} {\bibinfo  {journal} {Phys. Rev. E}\ }\textbf {\bibinfo {volume}
  {94}},\ \bibinfo {pages} {062410} (\bibinfo {year} {2016})}\BibitemShut
  {NoStop}%
\bibitem [{\citenamefont {Aguilar}\ \emph {et~al.}(2021)\citenamefont
  {Aguilar}, \citenamefont {Barbosa}, \citenamefont {Donangelo},\ and\
  \citenamefont {Souza}}]{Aguilar_PRE_2021_QuorumSensing}%
  \BibitemOpen
  \bibfield  {author} {\bibinfo {author} {\bibfnamefont {E.~J.}\ \bibnamefont
  {Aguilar}}, \bibinfo {author} {\bibfnamefont {V.~C.}\ \bibnamefont
  {Barbosa}}, \bibinfo {author} {\bibfnamefont {R.}~\bibnamefont {Donangelo}},
  \ and\ \bibinfo {author} {\bibfnamefont {S.~R.}\ \bibnamefont {Souza}},\
  }\href {\doibase 10.1103/PhysRevE.103.012403} {\bibfield  {journal} {\bibinfo
   {journal} {Phys. Rev. E}\ }\textbf {\bibinfo {volume} {103}},\ \bibinfo
  {pages} {012403} (\bibinfo {year} {2021})}\BibitemShut {NoStop}%
\bibitem [{\citenamefont {B\"auerle}\ \emph {et~al.}(2018)\citenamefont
  {B\"auerle}, \citenamefont {Fischer}, \citenamefont {Speck},\ and\
  \citenamefont {Bechinger}}]{Tobias_Nat_Commun_2018_QuorumSensing}%
  \BibitemOpen
  \bibfield  {author} {\bibinfo {author} {\bibfnamefont {T.}~\bibnamefont
  {B\"auerle}}, \bibinfo {author} {\bibfnamefont {A.}~\bibnamefont {Fischer}},
  \bibinfo {author} {\bibfnamefont {T.}~\bibnamefont {Speck}}, \ and\ \bibinfo
  {author} {\bibfnamefont {C.}~\bibnamefont {Bechinger}},\ }\href {\doibase
  10.1038/s41467-018-05675-7} {\bibfield  {journal} {\bibinfo  {journal} {Nat.
  Commun.}\ }\textbf {\bibinfo {volume} {9}},\ \bibinfo {pages} {3232}
  (\bibinfo {year} {2018})}\BibitemShut {NoStop}%
\bibitem [{\citenamefont {Zampetaki}\ \emph {et~al.}(2021)\citenamefont
  {Zampetaki}, \citenamefont {Liebchen}, \citenamefont {Ivlev},\ and\
  \citenamefont {L\"owen}}]{Lowen_PNAS_2021}%
  \BibitemOpen
  \bibfield  {author} {\bibinfo {author} {\bibfnamefont {A.~V.}\ \bibnamefont
  {Zampetaki}}, \bibinfo {author} {\bibfnamefont {B.}~\bibnamefont {Liebchen}},
  \bibinfo {author} {\bibfnamefont {A.~V.}\ \bibnamefont {Ivlev}}, \ and\
  \bibinfo {author} {\bibfnamefont {H.}~\bibnamefont {L\"owen}},\ }\href
  {\doibase 10.1073/pnas.2111142118} {\bibfield  {journal} {\bibinfo  {journal}
  {Proc. Natl. Acad. Sci. U.S.A.}\ }\textbf {\bibinfo {volume} {118}},\
  \bibinfo {pages} {e2111142118} (\bibinfo {year} {2021})}\BibitemShut
  {NoStop}%
\bibitem [{\citenamefont {Farkas}\ \emph {et~al.}(2003)\citenamefont {Farkas},
  \citenamefont {Helbing},\ and\ \citenamefont {Vicsek}}]{farkas_physA_2003}%
  \BibitemOpen
  \bibfield  {author} {\bibinfo {author} {\bibfnamefont {I.}~\bibnamefont
  {Farkas}}, \bibinfo {author} {\bibfnamefont {D.}~\bibnamefont {Helbing}}, \
  and\ \bibinfo {author} {\bibfnamefont {T.}~\bibnamefont {Vicsek}},\ }\href
  {\doibase 10.1016/j.physa.2003.08.014} {\bibfield  {journal} {\bibinfo
  {journal} {Physica A}\ }\textbf {\bibinfo {volume} {330}},\ \bibinfo {pages}
  {18} (\bibinfo {year} {2003})}\BibitemShut {NoStop}%
\bibitem [{\citenamefont {K{\"u}rsten}\ and\ \citenamefont
  {Ihle}(2020)}]{kursten_PRL_2020}%
  \BibitemOpen
  \bibfield  {author} {\bibinfo {author} {\bibfnamefont {R.}~\bibnamefont
  {K{\"u}rsten}}\ and\ \bibinfo {author} {\bibfnamefont {T.}~\bibnamefont
  {Ihle}},\ }\href {\doibase 10.1103/PhysRevLett.125.188003} {\bibfield
  {journal} {\bibinfo  {journal} {Phys. Rev. Lett.}\ }\textbf {\bibinfo
  {volume} {125}},\ \bibinfo {pages} {188003} (\bibinfo {year}
  {2020})}\BibitemShut {NoStop}%
\bibitem [{\citenamefont {Ginelli}\ \emph {et~al.}(2010)\citenamefont
  {Ginelli}, \citenamefont {Peruani}, \citenamefont {B{\"a}r},\ and\
  \citenamefont {Chat{\'e}}}]{ginelli_PRL_2010}%
  \BibitemOpen
  \bibfield  {author} {\bibinfo {author} {\bibfnamefont {F.}~\bibnamefont
  {Ginelli}}, \bibinfo {author} {\bibfnamefont {F.}~\bibnamefont {Peruani}},
  \bibinfo {author} {\bibfnamefont {M.}~\bibnamefont {B{\"a}r}}, \ and\
  \bibinfo {author} {\bibfnamefont {H.}~\bibnamefont {Chat{\'e}}},\ }\href
  {\doibase 10.1103/PhysRevLett.104.184502} {\bibfield  {journal} {\bibinfo
  {journal} {Phys. Rev. Lett.}\ }\textbf {\bibinfo {volume} {104}},\ \bibinfo
  {pages} {184502} (\bibinfo {year} {2010})}\BibitemShut {NoStop}%
\bibitem [{\citenamefont {Sumino}\ \emph {et~al.}(2012)\citenamefont {Sumino},
  \citenamefont {Nagai}, \citenamefont {Shitaka}, \citenamefont {Tanaka},
  \citenamefont {Yoshikawa}, \citenamefont {Chat{\'e}},\ and\ \citenamefont
  {Oiwa}}]{sumino_Nat_2012}%
  \BibitemOpen
  \bibfield  {author} {\bibinfo {author} {\bibfnamefont {Y.}~\bibnamefont
  {Sumino}}, \bibinfo {author} {\bibfnamefont {K.~H.}\ \bibnamefont {Nagai}},
  \bibinfo {author} {\bibfnamefont {Y.}~\bibnamefont {Shitaka}}, \bibinfo
  {author} {\bibfnamefont {D.}~\bibnamefont {Tanaka}}, \bibinfo {author}
  {\bibfnamefont {K.}~\bibnamefont {Yoshikawa}}, \bibinfo {author}
  {\bibfnamefont {H.}~\bibnamefont {Chat{\'e}}}, \ and\ \bibinfo {author}
  {\bibfnamefont {K.}~\bibnamefont {Oiwa}},\ }\href
  {https://www.nature.com/articles/nature10874} {\bibfield  {journal} {\bibinfo
   {journal} {Nature}\ }\textbf {\bibinfo {volume} {483}},\ \bibinfo {pages}
  {448} (\bibinfo {year} {2012})}\BibitemShut {NoStop}%
\bibitem [{\citenamefont {Carleo}\ \emph {et~al.}(2019)\citenamefont {Carleo},
  \citenamefont {Cirac}, \citenamefont {Cranmer}, \citenamefont {Daudet},
  \citenamefont {Schuld}, \citenamefont {Tishby}, \citenamefont
  {Vogt-Maranto},\ and\ \citenamefont {Zdeborov\'a}}]{Carleo_RMP_2019}%
  \BibitemOpen
  \bibfield  {author} {\bibinfo {author} {\bibfnamefont {G.}~\bibnamefont
  {Carleo}}, \bibinfo {author} {\bibfnamefont {I.}~\bibnamefont {Cirac}},
  \bibinfo {author} {\bibfnamefont {K.}~\bibnamefont {Cranmer}}, \bibinfo
  {author} {\bibfnamefont {L.}~\bibnamefont {Daudet}}, \bibinfo {author}
  {\bibfnamefont {M.}~\bibnamefont {Schuld}}, \bibinfo {author} {\bibfnamefont
  {N.}~\bibnamefont {Tishby}}, \bibinfo {author} {\bibfnamefont
  {L.}~\bibnamefont {Vogt-Maranto}}, \ and\ \bibinfo {author} {\bibfnamefont
  {L.}~\bibnamefont {Zdeborov\'a}},\ }\href {\doibase
  10.1103/RevModPhys.91.045002} {\bibfield  {journal} {\bibinfo  {journal}
  {Rev. Mod. Phys.}\ }\textbf {\bibinfo {volume} {91}},\ \bibinfo {pages}
  {045002} (\bibinfo {year} {2019})}\BibitemShut {NoStop}%
\bibitem [{\citenamefont {Carrasquilla}(2020)}]{Carrasquilla_AdvPhysX_2020}%
  \BibitemOpen
  \bibfield  {author} {\bibinfo {author} {\bibfnamefont {J.}~\bibnamefont
  {Carrasquilla}},\ }\href {\doibase 10.1080/23746149.2020.1797528} {\bibfield
  {journal} {\bibinfo  {journal} {Adv. Phys. X}\ }\textbf {\bibinfo {volume}
  {5}},\ \bibinfo {pages} {1797528} (\bibinfo {year} {2020})}\BibitemShut
  {NoStop}%
\bibitem [{\citenamefont {Cichos}\ \emph {et~al.}(2020)\citenamefont {Cichos},
  \citenamefont {Gustavsson}, \citenamefont {Mehlig},\ and\ \citenamefont
  {Volpe}}]{Volpe_Nat_Mach_Intell_2020}%
  \BibitemOpen
  \bibfield  {author} {\bibinfo {author} {\bibfnamefont {F.}~\bibnamefont
  {Cichos}}, \bibinfo {author} {\bibfnamefont {K.}~\bibnamefont {Gustavsson}},
  \bibinfo {author} {\bibfnamefont {B.}~\bibnamefont {Mehlig}}, \ and\ \bibinfo
  {author} {\bibfnamefont {G.}~\bibnamefont {Volpe}},\ }\href {\doibase
  10.1038/s42256-020-0146-9} {\bibfield  {journal} {\bibinfo  {journal} {Nat.
  Mach. Intell.}\ }\textbf {\bibinfo {volume} {2}},\ \bibinfo {pages} {94}
  (\bibinfo {year} {2020})}\BibitemShut {NoStop}%
\bibitem [{\citenamefont {Carrasquilla}\ and\ \citenamefont
  {Melko}(2017)}]{Melko_Nat_Phys_2017}%
  \BibitemOpen
  \bibfield  {author} {\bibinfo {author} {\bibfnamefont {J.}~\bibnamefont
  {Carrasquilla}}\ and\ \bibinfo {author} {\bibfnamefont {R.~G.}\ \bibnamefont
  {Melko}},\ }\href {\doibase 10.1038/nphys4035} {\bibfield  {journal}
  {\bibinfo  {journal} {Nat. Phys.}\ }\textbf {\bibinfo {volume} {13}},\
  \bibinfo {pages} {431} (\bibinfo {year} {2017})}\BibitemShut {NoStop}%
\bibitem [{\citenamefont {van Nieuwenburg}\ \emph {et~al.}(2017)\citenamefont
  {van Nieuwenburg}, \citenamefont {Liu},\ and\ \citenamefont
  {Huber}}]{van_Nieuwenburg_Nat_Phys_2017}%
  \BibitemOpen
  \bibfield  {author} {\bibinfo {author} {\bibfnamefont {E.~P.~L.}\
  \bibnamefont {van Nieuwenburg}}, \bibinfo {author} {\bibfnamefont {Y.-H.}\
  \bibnamefont {Liu}}, \ and\ \bibinfo {author} {\bibfnamefont {S.~D.}\
  \bibnamefont {Huber}},\ }\href {\doibase 10.1038/nphys4037} {\bibfield
  {journal} {\bibinfo  {journal} {Nat. Phys.}\ }\textbf {\bibinfo {volume}
  {13}},\ \bibinfo {pages} {435} (\bibinfo {year} {2017})}\BibitemShut
  {NoStop}%
\bibitem [{\citenamefont {Li}\ and\ \citenamefont
  {Wang}(2018)}]{Wanglei_PRL_2018_RG}%
  \BibitemOpen
  \bibfield  {author} {\bibinfo {author} {\bibfnamefont {S.-H.}\ \bibnamefont
  {Li}}\ and\ \bibinfo {author} {\bibfnamefont {L.}~\bibnamefont {Wang}},\
  }\href {\doibase 10.1103/PhysRevLett.121.260601} {\bibfield  {journal}
  {\bibinfo  {journal} {Phys. Rev. Lett.}\ }\textbf {\bibinfo {volume} {121}},\
  \bibinfo {pages} {260601} (\bibinfo {year} {2018})}\BibitemShut {NoStop}%
\bibitem [{\citenamefont {Guo}\ and\ \citenamefont {He}(2023)}]{Guo_NJP_2023}%
  \BibitemOpen
  \bibfield  {author} {\bibinfo {author} {\bibfnamefont {W.-C.}\ \bibnamefont
  {Guo}}\ and\ \bibinfo {author} {\bibfnamefont {L.}~\bibnamefont {He}},\
  }\href {\doibase 10.1088/1367-2630/acef4e} {\bibfield  {journal} {\bibinfo
  {journal} {New J. Phys.}\ }\textbf {\bibinfo {volume} {25}},\ \bibinfo
  {pages} {083037} (\bibinfo {year} {2023})}\BibitemShut {NoStop}%
\bibitem [{\citenamefont {Acebr{\'o}n}\ \emph {et~al.}(2005)\citenamefont
  {Acebr{\'o}n}, \citenamefont {Bonilla}, \citenamefont {Vicente},
  \citenamefont {Ritort},\ and\ \citenamefont
  {Spigler}}]{Acebrn_RevOfModernPhys_2005}%
  \BibitemOpen
  \bibfield  {author} {\bibinfo {author} {\bibfnamefont {J.~A.}\ \bibnamefont
  {Acebr{\'o}n}}, \bibinfo {author} {\bibfnamefont {L.~L.}\ \bibnamefont
  {Bonilla}}, \bibinfo {author} {\bibfnamefont {C.~J.~P.}\ \bibnamefont
  {Vicente}}, \bibinfo {author} {\bibfnamefont {F.}~\bibnamefont {Ritort}}, \
  and\ \bibinfo {author} {\bibfnamefont {R.}~\bibnamefont {Spigler}},\ }\href
  {\doibase 10.1103/RevModPhys.77.137} {\bibfield  {journal} {\bibinfo
  {journal} {Rev. Mod. Phys.}\ }\textbf {\bibinfo {volume} {77}},\ \bibinfo
  {pages} {137} (\bibinfo {year} {2005})}\BibitemShut {NoStop}%
\bibitem [{\citenamefont {Rodrigues}\ \emph {et~al.}(2016)\citenamefont
  {Rodrigues}, \citenamefont {Peron}, \citenamefont {Ji},\ and\ \citenamefont
  {Kurths}}]{rodrigues_PhysRep_2016}%
  \BibitemOpen
  \bibfield  {author} {\bibinfo {author} {\bibfnamefont {F.~A.}\ \bibnamefont
  {Rodrigues}}, \bibinfo {author} {\bibfnamefont {T.~K.~D.}\ \bibnamefont
  {Peron}}, \bibinfo {author} {\bibfnamefont {P.}~\bibnamefont {Ji}}, \ and\
  \bibinfo {author} {\bibfnamefont {J.}~\bibnamefont {Kurths}},\ }\href
  {\doibase 10.1016/j.physrep.2015.10.008} {\bibfield  {journal} {\bibinfo
  {journal} {Phys. Rep.}\ }\textbf {\bibinfo {volume} {610}},\ \bibinfo {pages}
  {1} (\bibinfo {year} {2016})}\BibitemShut {NoStop}%
\bibitem [{\citenamefont {Goodfellow}\ \emph {et~al.}(2016)\citenamefont
  {Goodfellow}, \citenamefont {Bengio},\ and\ \citenamefont
  {Courville}}]{Goodfellow_Book_2016}%
  \BibitemOpen
  \bibfield  {author} {\bibinfo {author} {\bibfnamefont {I.}~\bibnamefont
  {Goodfellow}}, \bibinfo {author} {\bibfnamefont {Y.}~\bibnamefont {Bengio}},
  \ and\ \bibinfo {author} {\bibfnamefont {A.}~\bibnamefont {Courville}},\
  }\href {http://www.deeplearningbook.org} {\emph {\bibinfo {title} {Deep
  Learning}}}\ (\bibinfo  {publisher} {MIT Press, Cambridge},\ \bibinfo {year}
  {2016})\BibitemShut {NoStop}%
\bibitem [{\citenamefont {Kingma}\ and\ \citenamefont
  {Ba}(2015)}]{Kingma_ICLR_2015}%
  \BibitemOpen
  \bibfield  {author} {\bibinfo {author} {\bibfnamefont {D.~P.}\ \bibnamefont
  {Kingma}}\ and\ \bibinfo {author} {\bibfnamefont {J.}~\bibnamefont {Ba}},\
  }in\ \href {http://arxiv.org/abs/1412.6980} {\emph {\bibinfo {booktitle}
  {Proceedings of the 3rd International Conference on Learning
  Representations}}}\ (\bibinfo  {publisher} {ICLR, San Diego},\ \bibinfo
  {year} {2015})\BibitemShut {NoStop}%
\bibitem [{\citenamefont {Vaswani}\ \emph {et~al.}(2017)\citenamefont
  {Vaswani}, \citenamefont {Shazeer}, \citenamefont {Parmar}, \citenamefont
  {Uszkoreit}, \citenamefont {Jones}, \citenamefont {Gomez}, \citenamefont
  {Kaiser},\ and\ \citenamefont {Polosukhin}}]{vaswani_Advances_2017}%
  \BibitemOpen
  \bibfield  {author} {\bibinfo {author} {\bibfnamefont {A.}~\bibnamefont
  {Vaswani}}, \bibinfo {author} {\bibfnamefont {N.}~\bibnamefont {Shazeer}},
  \bibinfo {author} {\bibfnamefont {N.}~\bibnamefont {Parmar}}, \bibinfo
  {author} {\bibfnamefont {J.}~\bibnamefont {Uszkoreit}}, \bibinfo {author}
  {\bibfnamefont {L.}~\bibnamefont {Jones}}, \bibinfo {author} {\bibfnamefont
  {A.~N.}\ \bibnamefont {Gomez}}, \bibinfo {author} {\bibfnamefont
  {{\L}.}~\bibnamefont {Kaiser}}, \ and\ \bibinfo {author} {\bibfnamefont
  {I.}~\bibnamefont {Polosukhin}},\ }in\ \href
  {https://papers.nips.cc/paper_files/paper/2017/hash/3f5ee243547dee91fbd053c1c4a845aa-Abstract.html}
  {\emph {\bibinfo {booktitle} {Proceedings of the Advances in Neural
  Information Processing Systems 30}}}\ (\bibinfo  {publisher} {NIPS, Long
  Beach, CA},\ \bibinfo {year} {2017})\BibitemShut {NoStop}%
\bibitem [{\citenamefont {Alkin}\ \emph {et~al.}()\citenamefont {Alkin},
  \citenamefont {F{\"u}rst}, \citenamefont {Schmid}, \citenamefont {Gruber},
  \citenamefont {Holzleitner},\ and\ \citenamefont
  {Brandstetter}}]{alkin_arxiv_2024}%
  \BibitemOpen
  \bibfield  {author} {\bibinfo {author} {\bibfnamefont {B.}~\bibnamefont
  {Alkin}}, \bibinfo {author} {\bibfnamefont {A.}~\bibnamefont {F{\"u}rst}},
  \bibinfo {author} {\bibfnamefont {S.}~\bibnamefont {Schmid}}, \bibinfo
  {author} {\bibfnamefont {L.}~\bibnamefont {Gruber}}, \bibinfo {author}
  {\bibfnamefont {M.}~\bibnamefont {Holzleitner}}, \ and\ \bibinfo {author}
  {\bibfnamefont {J.}~\bibnamefont {Brandstetter}},\ }\href@noop {} {\ }\Eprint
  {http://arxiv.org/abs/2402.12365} {arXiv:2402.12365} \BibitemShut {NoStop}%
\bibitem [{\citenamefont {Geneva}\ and\ \citenamefont
  {Zabaras}(2022)}]{geneva_NN_2022}%
  \BibitemOpen
  \bibfield  {author} {\bibinfo {author} {\bibfnamefont {N.}~\bibnamefont
  {Geneva}}\ and\ \bibinfo {author} {\bibfnamefont {N.}~\bibnamefont
  {Zabaras}},\ }\href {\doibase 10.1016/j.neunet.2021.11.022} {\bibfield
  {journal} {\bibinfo  {journal} {Neural Netw.}\ }\textbf {\bibinfo {volume}
  {146}},\ \bibinfo {pages} {272} (\bibinfo {year} {2022})}\BibitemShut
  {NoStop}%
\bibitem [{\citenamefont {Xu}\ \emph {et~al.}(2024)\citenamefont {Xu},
  \citenamefont {Zhou},\ and\ \citenamefont {Bian}}]{xu_Fluids_2024}%
  \BibitemOpen
  \bibfield  {author} {\bibinfo {author} {\bibfnamefont {B.}~\bibnamefont
  {Xu}}, \bibinfo {author} {\bibfnamefont {Y.}~\bibnamefont {Zhou}}, \ and\
  \bibinfo {author} {\bibfnamefont {X.}~\bibnamefont {Bian}},\ }\href
  {https://doi.org/10.1063/5.0188998} {\bibfield  {journal} {\bibinfo
  {journal} {Phys. Fluids}\ }\textbf {\bibinfo {volume} {36}} (\bibinfo {year}
  {2024})}\BibitemShut {NoStop}%
\bibitem [{\citenamefont {Wu}\ \emph {et~al.}()\citenamefont {Wu},
  \citenamefont {Luo}, \citenamefont {Wang}, \citenamefont {Wang},\ and\
  \citenamefont {Long}}]{wu_arxiv_2024}%
  \BibitemOpen
  \bibfield  {author} {\bibinfo {author} {\bibfnamefont {H.}~\bibnamefont
  {Wu}}, \bibinfo {author} {\bibfnamefont {H.}~\bibnamefont {Luo}}, \bibinfo
  {author} {\bibfnamefont {H.}~\bibnamefont {Wang}}, \bibinfo {author}
  {\bibfnamefont {J.}~\bibnamefont {Wang}}, \ and\ \bibinfo {author}
  {\bibfnamefont {M.}~\bibnamefont {Long}},\ }\href@noop {} {\ }\Eprint
  {http://arxiv.org/abs/2402.02366} {arXiv:2402.02366} \BibitemShut {NoStop}%
\bibitem [{\citenamefont {Sutskever}\ \emph {et~al.}(2014)\citenamefont
  {Sutskever}, \citenamefont {Vinyals},\ and\ \citenamefont
  {Le}}]{sutskever_Advances_2014}%
  \BibitemOpen
  \bibfield  {author} {\bibinfo {author} {\bibfnamefont {I.}~\bibnamefont
  {Sutskever}}, \bibinfo {author} {\bibfnamefont {O.}~\bibnamefont {Vinyals}},
  \ and\ \bibinfo {author} {\bibfnamefont {Q.~V.}\ \bibnamefont {Le}},\ }in\
  \href
  {https://papers.nips.cc/paper_files/paper/2014/hash/a14ac55a4f27472c5d894ec1c3c743d2-Abstract.html}
  {\emph {\bibinfo {booktitle} {Proceedings of the Advances in Neural
  Information Processing Systems 27}}}\ (\bibinfo  {publisher} {NIPS, Montreal,
  Canada},\ \bibinfo {year} {2014})\BibitemShut {NoStop}%
\bibitem [{\citenamefont {He}\ \emph {et~al.}(2015)\citenamefont {He},
  \citenamefont {Zhang}, \citenamefont {Ren},\ and\ \citenamefont
  {Sun}}]{he_IEEE_2015}%
  \BibitemOpen
  \bibfield  {author} {\bibinfo {author} {\bibfnamefont {K.}~\bibnamefont
  {He}}, \bibinfo {author} {\bibfnamefont {X.}~\bibnamefont {Zhang}}, \bibinfo
  {author} {\bibfnamefont {S.}~\bibnamefont {Ren}}, \ and\ \bibinfo {author}
  {\bibfnamefont {J.}~\bibnamefont {Sun}},\ }in\ \href
  {https://arxiv.org/abs/1502.01852} {\emph {\bibinfo {booktitle} {Proceedings
  of the IEEE International Conference on Computer Vision}}}\ (\bibinfo
  {publisher} {ICCV, Santiago, Chile},\ \bibinfo {year} {2015})\BibitemShut
  {NoStop}%
\end{thebibliography}

\end{document}